\documentclass[journal]{IEEEtran}
\IEEEoverridecommandlockouts

\usepackage{cite}
\usepackage{amsmath,amssymb,amsfonts}
\usepackage{graphicx}
\usepackage{textcomp}
\usepackage{xcolor}
\def\BibTeX{{\rm B\kern-.05em{\sc i\kern-.025em b}\kern-.08em
    T\kern-.1667em\lower.7ex\hbox{E}\kern-.125emX}}
\usepackage{pgfplots}
\pgfplotsset{compat=newest}
\usetikzlibrary{external}
    \usepackage{ragged2e}
\usepackage{floatrow}
\usepackage{tabu}
\usepackage{multirow}
\usepackage{amsthm}
\usepackage[utf8]{inputenc}
\usepackage[english]{babel}
\theoremstyle{remark}
\usepackage[export]{adjustbox}
\theoremstyle{plain} 

\newtheorem{proposition}{Proposition}

\newtheorem{remark}{Remark}

\newtheorem{corollary}{Corollary}
\usepackage{multicol}
\usepackage{epsfig}
\usepackage{epstopdf}
\usepackage{float}
\usepackage{perpage}
\MakeSorted{figure}
\MakeSorted{table}
\usepackage[normalem]{ulem}
\usepackage{array}
\usepackage{graphicx}
\usepackage{amsfonts}
\usepackage{amssymb}
\usepackage{soul}
\let\labelindent\relax
\usepackage{enumitem}							
\usepackage[english]{babel}
\usepackage[utf8]{inputenc}
\usepackage[linesnumbered,ruled,vlined]{algorithm2e}
\usepackage{upgreek}
\usepackage{bm} 
\usepackage{hyperref}
\usepackage{float}
\floatstyle{plaintop}
\restylefloat{table}
\hypersetup{linktocpage} 
\hypersetup{
colorlinks,
citecolor=black,
filecolor=black,
linkcolor=black,
urlcolor=black
}
\usepackage{tikz}
\usetikzlibrary{arrows.meta}
\usepackage{pifont}
\usepackage{makecell}
\usepackage{algorithmic}
\makeatletter
\renewenvironment{thebibliography}[1]{%
\@xp\section\@xp*\@xp{\refname}%
\normalfont\footnotesize\labelsep 1em\relax
\renewcommand\theenumiv{\arabic{enumiv}}\let\p@enumiv\@empty
\vspace*{0pt}
\list{\@biblabel{\theenumiv}}{\settowidth\labelwidth{\@biblabel{#1}}%
	\leftmargin\labelwidth \advance\leftmargin\labelsep
	\usecounter{enumiv}}%
\sloppy \clubpenalty\@M \widowpenalty\clubpenalty
\sfcode`\.=\@m
}{%
\def\@noitemerr{\@latex@warning{Empty `thebibliography' environment}}%
\endlist
}
\makeatother
\usepackage{pgfplots}

\usepackage{cite}
\DeclareMathAlphabet\mathbfcal{OMS}{cmsy}{b}{n}
\ifCLASSINFOpdf
\else
\fi
\usepackage{pgfplots}
\pgfplotsset{compat=newest}
\usepackage{tikz}

\usepackage{array}
\usepackage{fixltx2e}

\usepackage[font=footnotesize]{caption} 
\usepackage{subcaption}
\usepackage{fix-cm}
\usepackage{comment}
\usepackage{pgfplotstable}
\usepackage{pgfplots}
\usepgfplotslibrary{colormaps}
\usepgfplotslibrary{patchplots}

\usepackage{tikz}
\usepackage{pgfplots}
\pgfplotsset{compat=newest}
\usepgfplotslibrary{external}

\usepackage{mathtools}
\usepackage{pifont}
\newcommand{\norm}[1]{\left\lVert#1\right\rVert}

\usepackage{mathtools}
\usepackage{pifont}

\allowdisplaybreaks

\usepackage{tikz}

\title{POLO: Phase-Only Localization in Uplink Distributed MIMO Systems}
\author{Sauradeep Dey, Musa~Furkan~Keskin~\textit{Member, IEEE}, Dario~Tagliaferri,~\textit{Member, IEEE}, Gonzalo~Seco-Granados,~\textit{Fellow, IEEE},  Mikko~Valkama,~\textit{Fellow, IEEE}, Henk~Wymeersch, \textit{Fellow, IEEE}\vspace{-20pt}
\thanks{{This work was partly supported by the SNS JU project 6G-DISAC under the EU’s Horizon Europe research and innovation program under Grant Agreement No 101139130, the Swedish Research Council (VR) through the project 6G-PERCEF under Grant 2024-04390, the Spanish R+D project PID2023-152820OB-I00, and the AGAUR-ICREA Academia Program.}}
	\thanks{S. Dey, M. F. Keskin and H. Wymeersch are with the Department of Electrical Engineering, Chalmers University of Technology, Gothenburg, Sweden (emails: \{deysa, furkan, henkw\}@chalmers.se)
}	
\thanks{D. Tagliaferri is with the Department of Electronics, Information and Bioengineering, Politecnico di Milano, 20133, Milano, Italy (e-mail: dario.tagliaferri@polimi.it).}
\thanks{G. Seco-Granados is with the Department of Telecommunications and Systems Engineering, Universitat Autonoma de Barcelona, Spain (email:  gonzalo.seco@uab.cat).}
\thanks{ M. Valkama is with Tampere Wireless Research Center, Department of Electrical Engineering, Tampere University, 33100 Tampere, Finland (e-mail: mikko.valkama@tuni.fi)}
}
\begin{document}
\maketitle
\bstctlcite{IEEEexample:BSTcontrol}

\begin{abstract}
We propose a low-complexity localization framework for uplink distributed MIMO (D-MIMO) systems, targeting the challenge of minimizing the highly spiky maximum-likelihood (ML) cost function that arises in sparsely deployed phase-coherent access points (APs) with narrowband transmission. In such systems, ML-based localization typically relies on dense grid search, incurring prohibitive computational complexity.
To address this, we introduce phase-only localization (POLO), an approach that leverages differential carrier-phase measurements from selected APs to generate a compact set of candidate user positions. The ML cost function is then evaluated only at these candidates, reducing complexity significantly. A key challenge is to devise an AP selection mechanism that reduces the number of candidate points while maintaining reliable coverage. We propose two variants: POLO-I, which selects three APs to provide closed-form candidate positions with  low computational cost, and POLO-II, which selects four APs using an alternative strategy that enhances coverage at marginally higher runtime.
Comprehensive analytical and simulation results show that POLO achieves a favorable coverage-complexity trade-off, reducing cost by orders of magnitude relative to exhaustive grid search with only marginal loss in coverage. By characterizing this trade-off under diverse AP configurations, we also provide practical guidelines for selecting between POLO-I and POLO-II depending on latency and coverage requirements.
\end{abstract}
\vspace{-5mm}
\section{Introduction}
{Accurate localization of user equipment (UE) has become a key enabler in 5G and beyond-5G networks, supporting applications ranging from vehicular communication and industrial automation to augmented reality~\cite{LOC_SURVEY_1,6g_loc_enabler}. Besides existing systems that primarily rely on global navigation satellite systems (GNSS), next-generation wireless networks aim to provide high-precision, infrastructure-based localization by leveraging the large cellular communication infrastructure~\cite{5g_pos_ericsson,LOC_TUT_ABU}. 
Conventional localization techniques based on measurements such as time difference of arrival (TDoA) and angle of arrival (AoA) have been used in wireless networks. Their accuracy, however, is constrained by available system resources, namely the signal bandwidths and the aperture of the antenna arrays at base stations (BSs)\cite{LOC_SURVEY_1}.

\begin{figure}[t]
	\centering
		\includegraphics[width=1\linewidth]{Figures/pdf/System_model.pdf}\vspace{-7mm}
	\caption{Uplink carrier-phase measurement model in a D-MIMO system. 
    The UE at position $\mathbf{u}$ transmits to multiple APs located at $\mathbf{p}_m$. 
    The carrier-phase measurement at AP $m$ is $r_m$, with $\lambda$, $\phi$, and $n_m$ being the carrier wavelength, phase offset, and measurement noise, respectively. 
    The modulo-$2\pi$ operation wraps the phase onto the unit circle, resulting in integer ambiguity.}
	\label{fig_hyp}\vspace{0pt}
\end{figure}

{As 6G networks are expected to operate primarily in {FR1} and {FR3} frequency bands (where available bandwidth is limited), the focus is shifting toward leveraging large antenna arrays to achieve high-precision localization~\cite{LOC_SURVEY_5}.
Distributed MIMO (D-MIMO), in which multiple spatially distributed access points (APs) cooperate in a coordinated manner, has emerged as a promising solution for realizing such large-scale array deployments~\cite{NF_SURVEY_1}.
In particular, phase-coherent D-MIMO networks operate as large, sparsely distributed antenna array, where the UE propagation typically occurs in the near field~{\cite{NF_SURVEY_2}}.
This can be exploited to attain extremely high localization accuracy, aligning well with the stringent positioning requirements of 6G~\cite{6g_loc_sen_Hui_2024}.}
While TDoA- and AoA-based methods typically deliver up to sub-meter-level accuracy, D-MIMO systems can further improve accuracy (down to centimeter-level) by exploiting phase-coherence through \textit{carrier phase positioning (CPP)} techniques, which utilize the phase of the received observations to infer distances with sub-wavelength resolution~\cite{CPP_EMIL,CPP_FAN,CPP_HENK}. However, due to the periodic nature of the carrier wave, the measured phase is ambiguous by integer multiples of the wavelength \cite{CPP_TALVITIE}.
This \emph{integer ambiguity} makes positioning inherently challenging since there is not a one-to-one mapping between distance and observed~phase~\cite{CPP_MIKKO,CPP_HENK}.

\vspace{-7pt}
\subsection{{Related Works}}

CPP for UE localization through integer ambiguity resolution has been extensively investigated in recent studies~\cite{CPP_ZHANG,CPP_TALVITIE,CPP_FAN,CPP_JIN,CPP_KIM,CPP_FAN2,CPP_FOUDA,CPP_MIKKO}.
The authors in \cite{CPP_TALVITIE} and \cite{CPP_ZHANG} use high resolution carrier phase measurements for positioning by employing an extended Kalman filter (EKF) approach to resolve integer ambiguity over multiple snapshots, enabling centimeter-level accuracy. In addition, the work in \cite{CPP_FAN} proposes a carrier-phase-based fusion framework that integrates coarse UE position estimates from TDoA measurements with fine-grained updates from temporal changes in carrier phase. The fused estimates are then used to linearize the carrier-phase model to estimate the integer ambiguities. Following a carrier phase tracking approach with multiple snapshots, the work in \cite{CPP_JIN} introduces the double-difference carrier phase measurements method to address the integer ambiguity issue by continuously transmitting reference signals from two BSs and using additional measurements from a reference receiver. Similarly, the authors in \cite{CPP_FAN2} propose a factor
graph-based positioning method based on time-differential carrier phase measurements.
In \cite{CPP_FOUDA}, a dual-frequency virtual wavelength approach is proposed for UE localization, where two carriers at different frequencies are transmitted, and their phase difference is used to synthesize a larger virtual wavelength. While this method reduces the integer ambiguity search space, it suffers from increased resource consumption due to the need for multiple carrier transmissions.
A common drawback of the CPP methods in \cite{CPP_TALVITIE,CPP_ZHANG,CPP_FAN,CPP_JIN,CPP_FAN2} is that they require \textit{multiple snapshots} to resolve integer ambiguity, increasing latency in localization.
{Also, the works in \cite{CPP_ZHANG,CPP_FOUDA,CPP_JIN,CPP_KIM} leverage multiple {subcarriers} to improve ambiguity resolution, inherently increasing system resource usage.}
Another thead of research in CPP domain employs deep learning for localization \cite{CPP_MIKKO}. However, their effectiveness relies heavily~on the availability of abundant high-quality training~data, which can be cumbersome to collect for large~coverage~areas.

A promising alternative approach to solving the CPP problem is \textit{direct positioning}, which exploits the raw observations for \textit{single-snapshot} localization through maximum-likelihood estimation (MLE)~\cite{DOWNLINK_DIST,RS_UPLINK_JOURNAL,vukmirovic2019direct,DPE_GNSS_Magazine_2017,vukmirovic2018position}. Constructing the likelihood function directly from raw I/Q signals retains all phase and amplitude information, potentially improving estimation accuracy under low SNR and multipath propagation conditions \cite{DPE_GNSS_Magazine_2017}.
However, due to sparse deployment of spatially distributed phase-coherent APs in D-MIMO systems, the corresponding maximum likelihood (ML) cost function exhibits a highly non-convex and spiky landscape, making it difficult to reach the global optimum via standard optimization techniques~\cite{RS_UPLINK_JOURNAL,DOWNLINK_DIST}. Such a behavior stems from insufficient/sparse sampling of the spatial domain, which bears parallelism, for example, to sparse sampling of the time-frequency domain in OFDM radar, leading to ambiguities in delay-Doppler estimation (e.g., \cite{OFDM_DFRC_TSP_2021,Silvia_OFDM_sparse_TWC_2025}). Consequently, the MLE problem is generally solved numerically using dense grid search techniques, where the accuracy of the estimate is strongly influenced by the grid resolution~\cite{vukmirovic2019direct,RS_UPLINK_JOURNAL,DOWNLINK_DIST}. This, in turn, results in high computational complexity.
Moreover, the non-convexity of the likelihood function renders a potential gradient-based search highly sensitive to initialization, often requiring careful tuning or multi-start strategies to avoid getting trapped in spurious~peaks~\cite{vukmirovic2018position}.

\vspace{-10pt}
\subsection{Contributions}
{In this work, we propose phase-only localization (POLO), a novel framework for uplink phase-coherent D-MIMO systems based on \textit{single-snapshot} and narrowband/single-subcarrier observations at single-antenna APs. POLO leverages differential carrier-phase measurements together with an MLE cost function constructed directly from the raw received signals, tackling two major challenges in CPP: integer ambiguity and the non-convexity/spikiness of the MLE cost function. 
Unlike prior methods that resolve integer ambiguity by exploiting multiple snapshots (e.g. \cite{CPP_TALVITIE,CPP_ZHANG,CPP_FAN,CPP_JIN,CPP_FAN2}), {multiple subcarriers} (e.g. \cite{CPP_ZHANG,CPP_FOUDA,CPP_JIN,CPP_KIM,RS_UPLINK_JOURNAL}), or multi-antenna APs (e.g. \cite{RS_UPLINK_JOURNAL,DOWNLINK_DIST}), our approach achieves this using a single-snapshot, single-carrier, single-antenna setup. In doing so, POLO bridges the gap between the prohibitive complexity of MLE-based direct positioning methods \cite{RS_UPLINK_JOURNAL,DOWNLINK_DIST,vukmirovic2019direct,DPE_GNSS_Magazine_2017,vukmirovic2018position} and the resource-hungry integer ambiguity resolution strategies~\cite{CPP_ZHANG,CPP_FOUDA,CPP_JIN,CPP_KIM,RS_UPLINK_JOURNAL,CPP_TALVITIE,CPP_FAN2,DOWNLINK_DIST}.}
The main contributions of this paper are as follows:\vspace{-5pt}
\begin{enumerate}
\item {\textbf{Low-complexity phase-only localization for phase-coherent D-MIMO systems}}: 
We propose POLO, a low-complexity localization method for phase-coherent D-MIMO systems with single-carrier, single-snapshot transmission and single-antenna APs. 
POLO identifies a set of candidate UE locations that satisfy geometric constraints from differential phase measurements across AP pairs.
Instead of explicitly estimating the integer ambiguities, these candidate locations are used to initialize the MLE. This \textit{phase-guided grid pruning} drastically reduces the number of likelihood function evaluations, and therefore the computational cost as compared to the traditional exhaustive grid search (EGS), while maintaining high localization accuracy.

\item {\textbf{Novel AP subset selection strategies for performance-complexity tradeoff}}:
Two algorithmic POLO variants, distinguished by their AP selection strategy, are proposed: i) POLO-I, with two AP pairs sharing a common reference AP; and ii) POLO-II, with two distinct AP pairs. 
POLO-I allows closed-form UE candidate position solutions, offering  fast execution, suitable for systems with stringent time constraints. However, its localization coverage (fraction of the area with positioning error below a threshold) is limited by the geometry of the selected AP triplet.
POLO-II estimates the UE candidates in two steps: the first step evaluates raw candidate positions using two differential phase measurements from distinct AP pairs, and the second refines them by incorporating a third differential phase measurement. Although lacking a closed-form solution of UE candidates in the first step, and thus more computationally complex than POLO-I, POLO-II supports more flexible AP geometries and achieves improved localization coverage.

\item {\textbf{Geometry-aware performance analysis and AP selection}}: 
We perform detailed theoretical and simulation-based PEB analysis, revealing a strong spatial correlation between the PEB of the selected APs and the localization performance. Analytical derivations, verified by numerical simulations, identify specific regions in the coverage area, distinct for each POLO variant, where localization errors become unbounded. Based on these insights, we develop AP selection strategies that minimize high-error regions.
Further, we conduct extensive numerical evaluations across diverse AP configurations to provide practical guidelines for optimizing AP deployment and selection.
\end{enumerate}

\vspace{-10pt}
\section{System Model and Problem Description}
In this section, we present the system and signal model, and describe the problem for uplink D-MIMO localization.
\vspace{-10pt}
\subsection{System Model}\label{sec:system_model}
We consider a D-MIMO system operating in the uplink, where a single-antenna UE transmits a narrowband pilot symbol $s$, using a single-subcarrier at wavelength $\lambda$, to $M$ randomly deployed single-antenna APs. The pilot symbol is such that $|s|^2 = 1$. The 2D unknown position of the UE is denoted by $\mathbf{u} = [u_x, u_y]^\top$, while the known position of the $m$-th AP is $\mathbf{p}_m = [x_m, y_m]^\top$.
The APs are phase-synchronized, forming a virtual distributed array, and the UE has an unknown phase offset $\phi$ with respect to the AP network~\cite{RS_UPLINK_JOURNAL,DMIMO_FOR_6G,DOWNLINK_DIST}. 
The signal received by the $m$-th AP is given by\footnote{{{Similar to \cite{CPP_MIKKO,CPP_HENK, CPP_JIN,CPP_KIM, DOWNLINK_DIST,CPP_FAN2}}, we assume a fully line-of-sight (LoS) channel and neglect non-line-of-sight (NLoS) components.
This assumption emphasizes on the role of the direct path on the system performance. The resulting characterization can be interpreted as a best-case performance benchmark. The extensions to more realistic multipath environments can be considered in future works.}}
\begin{align}
y_{m} & =\sqrt{P}\rho_{m}e^{-\jmath\frac{2\pi }{\lambda} d_{m}} e^{- \jmath \phi} s + w_{m},\label{eq:signal}
\end{align}
where $P$ is the emitted power by the UE, $d_{m}=\Vert\mathbf{p}_{m}-\mathbf{u}\Vert$ denotes the distance between the $m$th AP and the UE, ${\rho_{m} = \sqrt{G_{\rm{tx}} G_{\rm{rx}}} \lambda/(4 \pi d_m)}$ represents the path-loss from the UE to the $m$-th AP, with $G_{\rm{tx}}$ and $G_{\rm{rx}}$ denoting the transmit and receive antenna gains, and $w_{m}\sim\mathcal{CN}(0,\sigma_w^2)$ is the additive white Gaussian noise (AWGN) at the $m$-th AP, uncorrelated across APs. From the received signal $y_m$, each AP obtains a measurement of the wrapped carrier phase as
\begin{align}
r_{m} = \angle y_m = -\frac{2\pi }{\lambda} d_m-\phi+2\pi z_{m}+n_{m},\label{eq:phasemodel}
\end{align}
where $r_m\in[0,2\pi)$, $z_m\in\mathbb{Z}$ denotes the integer ambiguity inherent in phase wrapping, and {$n_m$} is the is the noise on the carrier phase, whose power is inversely proportional to the signal-to-noise ratio (SNR)~\cite{SMKAY_BOOK_1}.

\subsection{Problem Description and Challenges}
Given {the received signal $\mathbf{y}=[y_1,\cdots,y_M]^\top$ in \eqref{eq:signal} or the phase measurements $\mathbf{r}=[r_1,\cdots,r_M]^\top$} in \eqref{eq:phasemodel}, the main objective of the system is to estimate the unknown UE position $\mathbf{u}$ in the presence of the unknown phase offset $\phi$ via a low-complexity approach. 
A common approach to localize the UE is to maximize the likelihood cost function constructed from the received signals $y_m$, as in~\cite{RS_UPLINK_JOURNAL,DOWNLINK_DIST}. However, due to the sparsely deployed nature of phase-coherent APs, the cost function is highly non-convex and contains multiple peaks, resulting in ambiguous location estimates and rendering standard optimization techniques ineffective. EGS can identify the correct position by evaluating the cost function over a dense grid, but this comes at the cost of high computational burden and strong sensitivity to grid resolution and initialization~{\cite{RS_UPLINK_JOURNAL,DOWNLINK_DIST,MLE_LEE,GRID_SEARCH_NAS,MLE_EGS}}.
Meanwhile, the phase measurements $r_m$ are observed modulo $2\pi$, introducing unknown integer ambiguities $z_m$. This leads to multiple candidate UE positions corresponding to different ambiguity values. 

\vspace{-10pt}
\subsection{{UE Localization Using MLE}}
We estimate the UE position $\mathbf{u}$, given the received signal $\mathbf{y}=[y_1,\cdots,y_M]^\top$ at the APs.
We frame the UE localization as an MLE problem as follows~\cite{SMKAY_BOOK_1}:}
\begin{align}\label{eq:prob1}
    &\{\widehat{\bm{\rho}}, \widehat{\phi}, \widehat{\mathbf{u}}\}=\underset{\bm{\rho}, \phi, \mathbf{u}}{ \arg\min }\;\; C(\bm{\rho}, \phi, \mathbf{u}), \text{ where}\\
    &C(\bm{\rho}, \phi, \mathbf{u}) = \sum\limits_{m=1}^M \Big| y_m - \sqrt{P}\rho_{m}e^{-\jmath\frac{2\pi }{\lambda} d_{m}(\mathbf{u})} e^{- \jmath \phi} s \Big|^2.\label{eq:cost1}
\end{align}
The unknown parameters include the channel amplitudes $\bm{\rho}=[\rho_1,\cdots,\rho_M]^\top$, the UE phase offset $\phi$, and its position $\mathbf{u}$. Note that the summation spans all the APs.
To express the cost function only in terms of $\mathbf{u}$, we {compress away} the unknowns $\rho_m$ and $\phi$~\cite{RS_UPLINK_JOURNAL,PAR_EST_BOOK,SMKAY_BOOK_1}. In particular, differentiating the cost function with respect to $\rho_m$ and equating it to zero, we obtain:
\begin{align}\label{eq:rho}
    \widehat{\rho}_m = \frac{1}{\sqrt{P}} \Re\Big\{ y_m^* e^{-\jmath\frac{2\pi }{\lambda} d_{m}(\mathbf{u})} e^{- \jmath \phi}s \Big\}.
\end{align}
Note that $\widehat{\rho}_m$ is a function of both $\phi$ and $\mathbf{u}$. Substituting $\widehat{\rho}_m$ into the cost function in \eqref{eq:cost1}, we obtain
\begin{align}
&C(\mathbf{u},\phi) = \sum\limits_{m=1}^M \left| y_m - \sqrt{P}\widehat{\rho}_{m}e^{-\jmath\frac{2\pi }{\lambda} d_{m}(\mathbf{u})} e^{- \jmath \phi} s \right|^2
\nonumber \\
&= \frac{1}{2} \sum\limits_{m=1}^M |y_m|^2 - \Re \Big\{ e^{j2\phi} \sum\limits_{m=1}^M (y_m^*)^2 e^{-j \frac{4\pi }{\lambda} d_m(\mathbf{u})} s \Big\}.
\label{eq:cost_2}
\end{align}
The value of $\phi$ that minimizes $C(\mathbf{u},\phi)$ maximizes the term within the real part, and it is given by
\begin{align}\label{eq:delta_phi}
    \widehat{\phi}(\mathbf{u}) = -\frac{1}{2} \angle \Big( \sum\limits_{m=1}^M (y_m^*)^2 e^{-j \frac{4\pi }{\lambda} d_m(\mathbf{u})} s \Big).
\end{align}
Substituting $\widehat{\phi}(\mathbf{u})$ from \eqref{eq:delta_phi}  in \eqref{eq:cost_2} and dropping the constants, we obtain the final cost function, which solely depends on~$\mathbf{u}$:
\begin{align}\label{eq:cost_final}
    C(\mathbf{u}) =  - \Re \Big\{ e^{j2\widehat{\phi}(\mathbf{u})} \sum\limits_{m=1}^M (y_m^*)^2 e^{-j \frac{4\pi }{\lambda} d_m(\mathbf{u})} s \Big\}.
\end{align}
The localization problem now becomes
\begin{align}\label{eq:opt_prob}
    &\widehat{\mathbf{u}}=\underset{\mathbf{u}}{ \arg\min }\;\; C(\mathbf{u}).
  \end{align}
The ML cost function $C(\mathbf{u})$ is highly non-convex, rendering standard optimization techniques ineffective~\cite{RS_UPLINK_JOURNAL,DOWNLINK_DIST}.
One way to address this is to evaluate $C(\mathbf{u})$ over a predefined set of candidate UE positions $\mathcal{U}$, and select the position with the minimum cost,
\begin{align}\label{eq:min_cost}
\widehat{\mathbf{u}}^{(0)} = \arg\min_{\mathbf{u} \in \mathcal{U}} C(\mathbf{u}).    
\end{align}
This initial estimate $\widehat{\mathbf{u}}^{(0)}$ is then refined by applying gradient descent on $C(\mathbf{u})$, yielding the final UE position~\cite{BOYD_CV_OPT},
\begin{align}\label{eq:gd}
    \widehat{\mathbf{u}} = \mathrm{GD}\big(C(\mathbf{u}), \widehat{\mathbf{u}}^{(0)}\big),
\end{align}
where $\mathrm{GD}(f(\mathbf{x}), \mathbf{x}^{(0)})$ denotes the gradient descent procedure applied to \(f(\mathbf{x})\) with \(\mathbf{x}^{(0)}\) as the initialization point~\cite{BOYD_CV_OPT}. 
{The non-convexity of $C(\mathbf{u})$ makes MLE highly sensitive to the initialization~\cite{RS_UPLINK_JOURNAL}. 
Next, we discuss the POLO framework, which instead of evaluating the likelihood function over a dense grid, leverages geometric constraints from the differential phase measurements to prune the candidate UE position set $\mathcal{U}$.} 

\vspace{-10pt}
\section{POLO-I:  Phase-Based Candidate Reduction for ML Localization}\label{sect:two-stage}
This section introduces POLO-I, a localization strategy that reduces the number of likelihood function evaluations by exploiting the periodic dependence of wrapped phase measurements $r_m$ in \eqref{eq:phasemodel} on the UE position $\mathbf{u}$ (via $d_m$). Specifically, three APs are selected: one reference ($\text{AP}_r$) and two secondary APs ($\text{AP}_{s_1}$, $\text{AP}_{s_2}$). Differential phase measurements between the reference AP and each secondary AP are exploited to generate a finite set of candidate UE positions. The likelihood cost function $C(\mathbf{u})$ is then evaluated only over this set, and the best candidate is refined via gradient descent. 
{We begin by deriving the candidate UE positions for a given AP triplet, and then propose two strategies for selecting the AP triplets.}

\subsection{Identification of Candidate UE Positions}\label{subsect:stage1}
We notice from \eqref{eq:phasemodel} that the wrapped phase $r_m$, assuming a fixed integer ambiguity $z_m$, depends on three unknowns: the 2D UE position $\mathbf{u}$ and the phase offset $\phi$. Thus, phase observations from at least three APs are required to identify a candidate UE position for a given value of the ambiguity.

We form a triplet $\{\text{AP}_r, \text{AP}_{s_1}, \text{AP}_{s_2}\}$, and compute the differential phases between the AP pairs $(\text{AP}_r, \text{AP}_{s_1})$ and $(\text{AP}_r, \text{AP}_{s_2})$ as~\cite{HYP_SOL_KCHO,TDOA_ANALYT_SOL}
\begin{align}
    \Delta r_{s_1} = \frac{\lambda}{2\pi}(r_r - r_{s_1}) 
    = (d_r - d_{s_1}) + \widetilde{z}_{s_1} \lambda + \widetilde{n}_{s_1} \label{eq:phase_diff_1}\\
    \Delta r_{s_2} = \frac{\lambda}{2\pi}(r_r - r_{s_2}) 
    = (d_r - d_{s_2}) + \widetilde{z}_{s_2} \lambda + \widetilde{n}_{s_2},\label{eq:phase_diff_2}
\end{align}
where $d_r$, $d_{s_1}$, and $d_{s_2}$ are the distances from the UE to the APs $\text{AP}_r$, $\text{AP}_{s_1}$, and $\text{AP}_{s_2}$, respectively. The integers $\widetilde{z}_{s_1} = z_{s_1} - z_r$ and $\widetilde{z}_{s_2} = z_{s_2} - z_r$  are the differential ambiguity, while $\widetilde{n}_{s_1} = {\lambda}(n_r - n_{s_1})/{2\pi}$ and $\widetilde{n}_{s_2} = {\lambda}( n_r - n_{s_2})/{2\pi}$ are the differential noise samples. Eqs. \eqref{eq:phase_diff_1}-\eqref{eq:phase_diff_2} represent two families of hyperbolas parameterized by the differential ambiguities $\widetilde{z}_{s_1}$ and $\widetilde{z}_{s_2}$, which for known AP positions lie within ranges $\mathcal{Z}_{s_1} = [Z_{s_1,\min}, Z_{s_1,\max}]$ and $\mathcal{Z}_{s_2} \!=\! [Z_{s_2,\min}, Z_{s_2,\max}]$:  
Given $(\widetilde{z}_{s_1}, \widetilde{z}_{s_2}) \in \mathcal{Z}_{s_1} \times \mathcal{Z}_{s_2}$, these hyperbolas are given by
\begin{align}
\mathcal{H}_1 &: \|\mathbf{p}_{s_1} - \mathbf{u}\| - \|\mathbf{p}_{r} - \mathbf{u}\| = \Delta r_{s_1} - \widetilde{z}_{s_1} \lambda,\label{eq:hyp_1} \\
\mathcal{H}_2 &: \|\mathbf{p}_{s_2} - \mathbf{u}\| - \|\mathbf{p}_{r} - \mathbf{u}\| = \Delta r_{s_2} - \widetilde{z}_{s_2} \lambda.\label{eq:hyp_2}
\end{align}
Their intersections define the set of candidate UE positions $\mathcal{U}$~\cite{HYP_SOL_KCHO}. Fig.~\ref{fig_hyp} illustrates the set of hyperbolas generated for a given AP triplet.  
 These intersections can be computed in closed form (see Appendix~\ref{app_hyp_intersection}). Finally, with the generated $\mathcal{U}$ we solve \eqref{eq:min_cost} and \eqref{eq:gd} to estimate the UE position.

We note that the number of generated hyperbola pairs, $N_{\text{P1}}$, is the cardinality of the Cartesian product set $\mathcal{Z}_{s_1} \times \mathcal{Z}_{s_2}$. 
It is readily verified that\footnote{{The differential phase measurements induce hyperbolas spaced {$\lambda/2$} apart between the reference AP and each secondary AP. For known $\mathbf{p}_r$ and $\mathbf{p}_{s_i}$ with $i\in\{1,2\}$ the differential ambiguity $\widetilde{z}_{s_i}$ lies between  $Z_{s_i,\min}= -\left\lfloor{\|\mathbf{p}_r-\mathbf{p}_{s_i}\|}/{\lambda}\right\rceil$ and $Z_{s_i,\max}= \left\lfloor{\|\mathbf{p}_r-\mathbf{p}_{s_i}\|}/{\lambda}\right\rceil$.}}
\begin{align}\label{eq:NP1}
    N_{\text{P1}} = \frac{4d_{s_1}  d_{s_2}}{\lambda^2}.
\end{align}
Each hyperbola pair yields {up to} two candidate UE positions (see Appendix A). Therefore, the candidate set $\mathcal{U}$ contains {at most} $|\mathcal{U}| = 2N_{\text{P1}}$ possible positions. 


\begin{figure}[t]
	\centering
	\begin{subfigure}[b]{0.8\linewidth}
		\includegraphics[width=0.9\linewidth,height=0.75\linewidth]{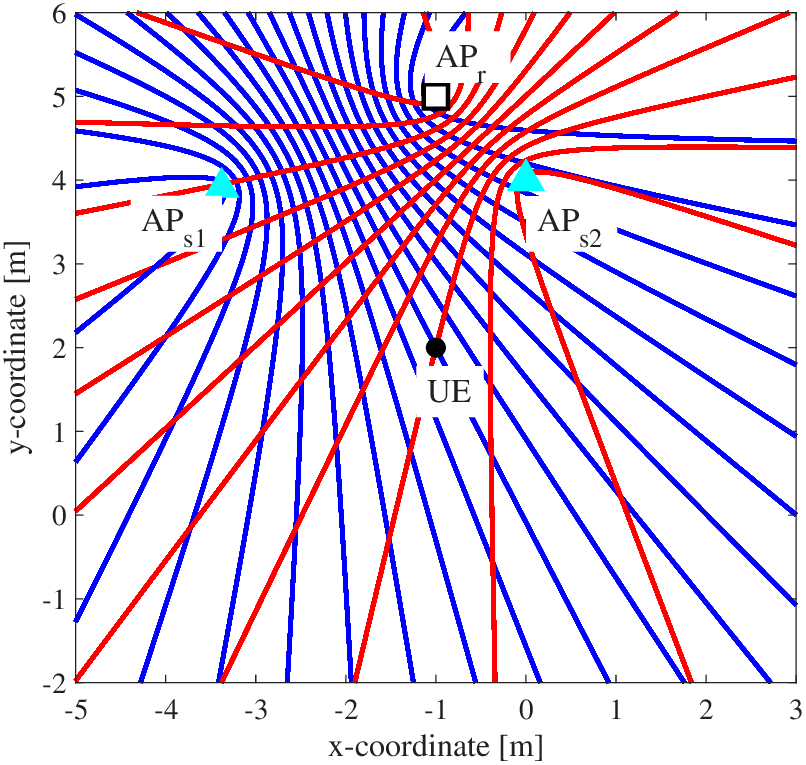}\vspace{0pt}
	\end{subfigure}
	\caption{{Illustration of hyperbolas generated from differential phase measurements between the reference AP (white square) and two secondary APs (blue triangles). 
    {The set of blue and red hyperbolas are generated using the AP pair $\{\text{AP}_r,\text{AP}_{s1}\}$, and  $\{\text{AP}_r,\text{AP}_{s2}\}$, respectively. Each hyperbola stems from different values of the differential integer ambiguities  $\widetilde{z}_{s_1}$ and $\widetilde{z}_{s_2}$.}
    The UE location is marked by the black dot. 
    }}
	\label{fig_hyp}\vspace{0pt}
\end{figure}

\subsection{AP Selection Strategies for POLO-I}\label{sec_AP_sel}
We now propose two different AP selection strategies for POLO-I. The performance of the proposed algorithm in terms of localization coverage and computational complexity are heavily dependent on the selection of AP triplet. Indeed, both the number and geometry of the generated hyperbolas depend on the spatial configuration of the~selected~APs. {Here, localization coverage refers to the fraction of the area where the localization error is below a predefined threshold.}

\subsubsection{Strategy 1: Low-Complexity AP Selection}
The first AP selection strategy focuses on minimizing computational complexity, i.e., reducing the number of candidate UE positions $|\mathcal{U}|$. 
{From \eqref{eq:NP1}, this leads to the following selection problem:
\begin{equation}
    {(\widehat{r},\widehat{s}_1,\widehat{s}_2) = \underset{r,s_1,s_2 }{\arg \min} \;\;\|\mathbf{p}_r-\mathbf{p}_{s_1}\| \cdot\|\mathbf{p}_r-\mathbf{p}_{s_2}\|},
\end{equation}
where $r,s_1,s_2 \in \{1,..,M\}$ denote the indices of $\text{AP}_r$, $\text{AP}_{s_1}$ and $ \text{AP}_{s_2}$, respectively.
}
Despite its computational efficiency, this strategy does not account for localization accuracy in the coverage area. The next strategy addresses this limitation.

\begin{figure}[t]
	\centering
	\begin{subfigure}[b]{\linewidth}\centering
		\includegraphics[width=\linewidth,height=0.8\linewidth]{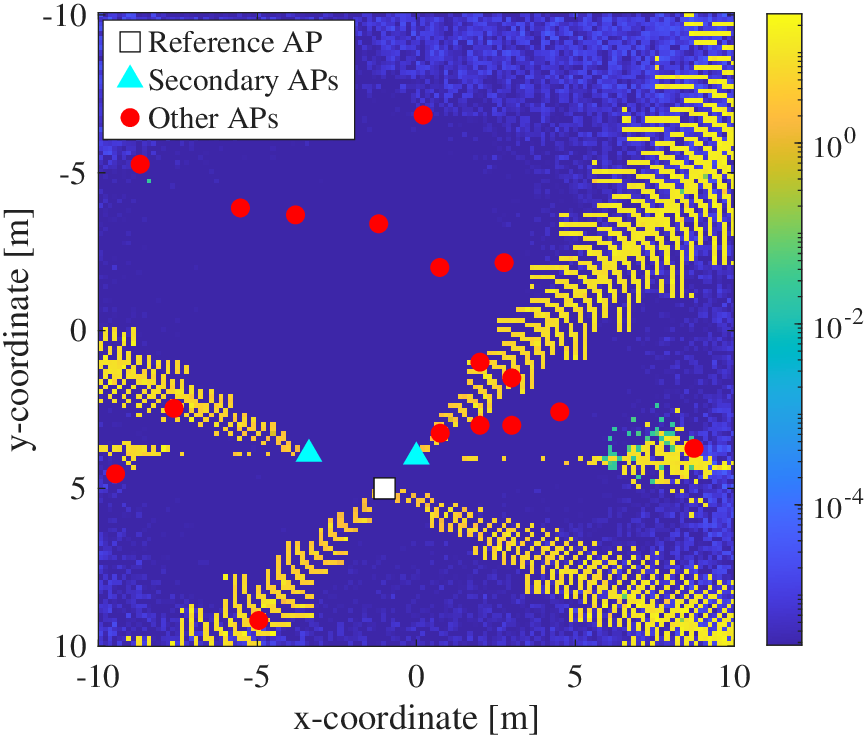}\vspace{0pt}
        \caption{ }  \label{fig_rmse_sample}
	\end{subfigure}
    \begin{subfigure}[b]{\linewidth}\centering
		\includegraphics[width=\linewidth,height=0.8\linewidth]{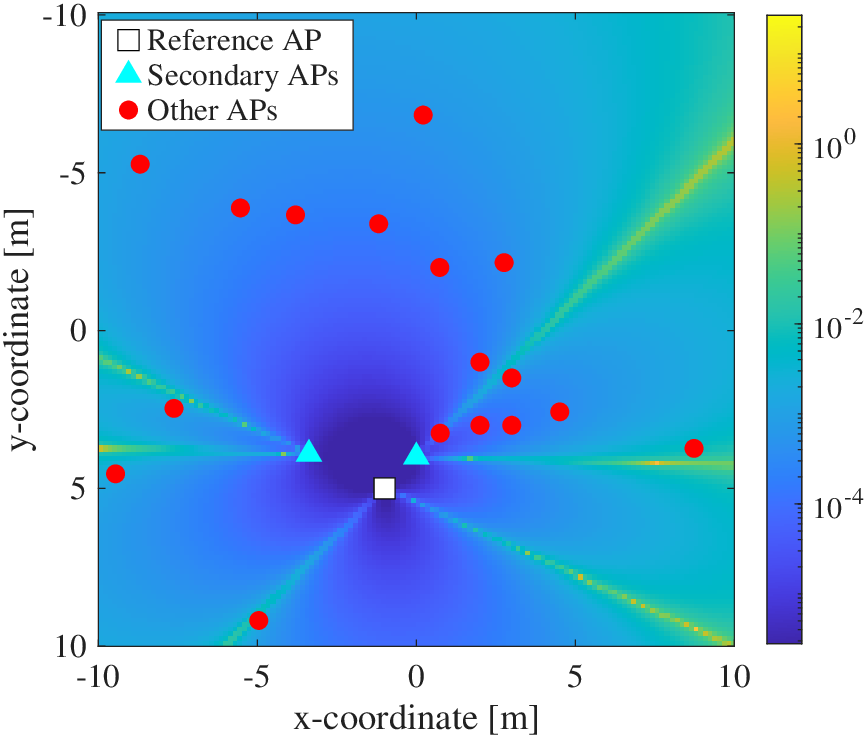}\vspace{0pt}
        \caption{ }  \label{fig_PEB_sample}
    \end{subfigure}
	\caption{{ a) RMSE map [in m]; and b) PEB map [in m] for different UE locations in the coverage area. The RMSE map in (a) is evaluated using the measurements from all $M$ APs, with the UE candidate set $\mathcal{U}$ generated using initially selected 3 APs. The PEB map in (b) is evaluated considering only the selected 3 APs. A strong spatial correlation between the RMSE and PEB maps is observed. Simulation parameters are provided in Section~\ref{sec_sim_scenario}.}}  
	\label{fig_sample}\vspace{0pt}
\end{figure}

\subsubsection{{Strategy 2: Position Error Bound (PEB)-Induced AP Selection}}\label{sec_AP_sel_2}
{The second strategy extends the first by limiting complexity while considering the localization coverage. To understand the location coverage, we first derive the PEB considering only the selected AP triplet $(r,s_1,s_2)$. 
\begin{proposition}\label{prop_EFIM}
Given selected triplet $\mathcal{M}=\{r,s_1,s_2\}$, the PEB for the UE position $\mathbf{u}$  is given by $\text{PEB}_T=\sqrt{\text{trace}(\mathbf{J}^{-1}_{T}{(\mathbf{u})}) }$, where $\mathbf{J}_{T}{(\mathbf{u})} $ is the  equivalent Fisher information matrix (EFIM)  \cite{EFIM_BOOK}, with
\begin{align}\label{eq:EFIM_sel_final}
    \mathbf{J}_{T}{(\mathbf{u})} = K_c\sum_{m\in\mathcal{M}}\rho_m^2\mathbf{v}_m\Big(\beta_m \mathbf{v}_m^\top
    -\!\!\!\!\sum_{m'\in\mathcal{M}\backslash m}\rho_{m'}^2\mathbf{v}_{m'}^\top\Big),\!\!\!\!
\end{align}
with  $K_c= {8P\pi^2}/{\Big(\sigma_n^2\lambda^2\sum\limits_{m'\in \mathcal{M}}\rho_{m'}^2\Big)}$, $\beta_m=\sum\limits_{m'\in\mathcal{M}\backslash m}\rho_{m'}^2$, and $\mathbf{v}_m={(\mathbf{p}_m-\mathbf{u})}/{d_m}$ is the unit vector from the UE to the $m$th AP.  The value of $\text{PEB}_T$ does not depend on the choice of reference AP among the selected triplet $\mathcal{M}$. 
\end{proposition}
\begin{proof}
    The proof is given in Appendix~\ref{app:PEB}.
\end{proof}
{We now analyze the correlation between the localization \textit{root mean square error (RMSE)} and the \textit{selected AP triplet PEB} $(\mathrm{PEB}_T)$. The RMSE represents the localization error of the UE position estimated from \eqref{eq:min_cost} and \eqref{eq:gd}. Since the candidate set $\mathcal{U}$ in \eqref{eq:min_cost} is generated solely from the differential phase measurements of the selected AP triplet, the quality of the initial estimate $\widehat{\mathbf{u}}^{(0)}$ is directly linked to $\mathrm{PEB}_T$. This establishes a clear connection between $\mathrm{PEB}_T$ and the overall localization RMSE across the coverage area.}

{To illustrate this relationship, Fig.~\ref{fig_sample} compares the RMSE of estimated UE positions (Fig.~\ref{fig_rmse_sample}) with $\mathrm{PEB}_T$ (Fig.~\ref{fig_PEB_sample}) for a specific AP triplet over the coverage area. The RMSE values are obtained by solving \eqref{eq:min_cost}, while $\mathrm{PEB}_T$ is computed using only the selected triplet. As expected, the RMSE and $\mathrm{PEB}_T$ maps exhibit a strong spatial correlation. 
We observe regions with high $\mathrm{PEB}_T$, radiating out from the AP triplet. These regions can be characterized exactly as follows. }
\vspace{-5pt}
\begin{corollary}\label{cor_EFIM}
Suppose the UE lies on the line defined by two selected APs, ${m_1,m_2}\in\mathcal{M}$, but {outside the segment joining them}, and the third AP $m_3$ is at an arbitrary location, i.e., $\mathbf{v}_{m_1}=\mathbf{v}_{m_2}=\mathbf{v}$ and  $\mathbf{v}_{m_3}\neq \mathbf{v}$. Then,  the EFIM becomes
\begin{align}\label{eq:EFIM_sp_case_POLO_I}
    \mathbf{J}_{T}(\mathbf{u})
    &= K_c\beta_{m_3}(\mathbf{v}-\mathbf{v}_{m_3})(\mathbf{v}-\mathbf{v}_{m_3})^\top,
\end{align}
leading to $\text{PEB}_T \to + \infty$.
\end{corollary}
\begin{proof}
When the UE lies on the straight line that passes through the APs $m_1$ and $m_2$, but outside the segment between these two APs, then $\mathbf{v}_{m_1}=\mathbf{v}_{m_2}=\mathbf{v}$, while $\mathbf{v}_{m_3}\neq \mathbf{v}$. Substituting this into \eqref{eq:EFIM_sel_final} directly yields \eqref{eq:EFIM_sp_case_POLO_I}.  Since the EFIM in this case is rank-deficient,  $\mathrm{PEB}_T$ becomes infinite.
\end{proof}
Notably, this degeneracy occurs only when the UE is \emph{outside} the segment connecting the two APs. When the UE lies within the segment, i.e., $\mathbf{v}_{m_1}=-\mathbf{v}_{m_2}$, the EFIM is full rank,  and $\mathrm{PEB}_T$ is finite. This explains the high-error regions in the RMSE and $\mathrm{PEB}_T$ maps of Fig.~\ref{fig_sample}.
{A similar observation was reported in \cite[Fig. 5]{TDOA_ANALYT_SOL}, where TDOA analysis showed  increased localization errors along lines joining sensors, although no formal mathematical explanation was provided.}

The observed correlation between the RMSE and $\mathrm{PEB}_T$ motivates a strategy that uses $\mathrm{PEB}_T$ as a  proxy performance metric. 
Formally, the AP selection strategy is formulated as
\begin{align}\label{eq:ap_trip_opt_1}
 (\widehat{r},\widehat{s_1},\widehat{s_2}) = \underset{(r,s_1,s_2)\in\mathcal{T} }{\arg \max}\;\;{f_{\mathrm{PEB}_T<\Gamma} \big(r,s_1,s_2\big)},    
\end{align}
where {the objective $f_{\mathrm{PEB}_T<\Gamma} \big(r,s_1,s_2\big)$, referred to as localization coverage, is the fraction of coverage area where the selected triplet PEB $\mathrm{PEB}_T$ is below a threshold $\Gamma$.} The set $\mathcal{T}$ includes all possible AP triplets with cardinality $\binom{M}{3}$. 
Since the number of possible triplets grows rapidly with~$M$, exhaustive evaluation of $\mathrm{PEB}_T$ for all AP triplets across the coverage area is computationally expensive. To reduce complexity, we restrict the search to a smaller set $\mathcal{T}_{\text{sel}} \subset \mathcal{T}$, whose construction is guided by the  theoretical analysis~of~$\mathrm{PEB}_T$. 

This smaller set of candidates $\mathcal{T}_{\text{sel}}$ is guided by Corollary \ref{cor_EFIM}: 
  {non-collinear AP triplets generate three distinct high-error lines between each AP pair, enlarging the area of poor localization. Collinear or nearly collinear triplets, however, confine this region to a single line. Further, triplets with shorter reference-to-secondary distances reduce computational complexity. Thus, choosing collinear or nearly collinear triplets with short inter-AP distances improves localization coverage while keeping the complexity low.} 
Drawing insights from the above analysis, we retain only those triplets that are:
\begin{itemize}
    \item \textit{Collinear or nearly collinear}, so that the high error region is limited to a single line, and
    \item \textit{Distance-bounded}, i.e., the reference–secondary AP separation is kept below a threshold to limit complexity.
\end{itemize}

 This filtering step produces a reduced candidate set $\mathcal{T}_{\text{sel}}$, with a cardinality significantly smaller than that of $\mathcal{T}$. The detailed selection procedure is provided in Algorithm~\ref{alg:ap_selection}.

\begin{algorithm}[t]
\caption{{POLO-I: AP Selection Strategy 2}}
\label{alg:ap_selection}
\begin{algorithmic}[1]
\STATE \textbf{Input:} Set of all AP positions $\{\mathbf{p}_m\}_{m=1}^M$, $\varepsilon=$ angular tolerance, $\gamma=$ AP distance threshold, $\lambda=$ wavelength, $\Gamma=$ $\mathrm{PEB}_T$ threshold, and $\mathcal{A}=$ Area of interest.
\STATE \textbf{Output:} Selected AP triplet $(\widehat{r}, \widehat{s}_1, \widehat{s}_2)$
\vspace{0.5em}
\STATE Define set of all AP triplets $\mathcal{T}=\{(m_1,m_2,m_3)|1\leq m_1<m_2<m_3\leq M\}$
\FOR{each $(m_1, m_2, m_3)\in\mathcal{T}$ and for each permutation of AP assignments $(r, s_1, s_2)$}
    \STATE Compute angle $\theta=\cos^{-1}\Big(\frac{(\mathbf{p}_{s_1} - \mathbf{p}_{r})^\top(\mathbf{p}_{s_2} - \mathbf{p}_{r})}{\norm{\mathbf{p}_{s_1} - \mathbf{p}_{r}}\norm{\mathbf{p}_{s_2} - \mathbf{p}_{r}}}\Big)$
    \IF{$|\theta - 180^\circ| \leq \varepsilon$}
        \STATE Classify the triplet as collinear  $(r,s_1,s_2)\in \mathcal{T}_\text{col}$
     \ENDIF
\ENDFOR
\vspace{0.5em}
\STATE For each $(r,s_1,s_2)\in\mathcal{T}_\text{col}$ compute $d_{rs_1s_2}= \max \{\| \mathbf{p}_{r} - \mathbf{p}_{s_1} \|,\| \mathbf{p}_{r} - \mathbf{p}_{s_2} \|,\| \mathbf{p}_{s_1} - \mathbf{p}_{s_2} \|\}$
\STATE Define a filtered set:  $\mathcal{T}_\text{sel}=\{ (r,s_1,s_2)\in\mathcal{T}_\text{col} | d_{rs_1s_2} < \gamma\}$.
\FOR{each $(r,s_1,s_2)\in\mathcal{T}_\text{sel}$}
    \STATE For each $(r,s_1,s_2) \in \mathcal{T}_{\text{sel}}$, evaluate $\mathrm{PEB}_T$ map $\mathrm{PEB}_T(\mathbf{u},r,s_1,s_2)$ over the UE position grid $\mathbf{u} \in \mathcal{A}$.
\ENDFOR
\STATE \textbf{Return} $(\widehat{r},\widehat{s_1},\widehat{s_2}) = \underset{(r,s_1,s_2)\in\mathcal{T}_\text{sel} }{\arg \min}\;\;{f_{\mathrm{PEB}_T<\Gamma} \big(r,s_1,s_2\big)}$.
\end{algorithmic}
\end{algorithm}

\subsection{{Limitations}} \label{sec_POLO_I_lim}
{POLO-I has a key limitation: its localization performance is reliable within the segment connecting the selected APs, but along the line extensions beyond the selected APs, the error becomes unbounded. Even with the proposed AP selection strategies, the high-error regions can span large areas limiting localization coverage. 
One way to improve coverage is by increasing the spacing between the selected APs. 
However, this also increases the number of hyperbola intersections, which increases the number of ML cost evaluations.
{Thus, in POLO-I the localization coverage is directly linked with the number of ML cost evaluations.}
To overcome this limitation, we introduce a variant of the algorithm, POLO-II, in the next section.}


\section{POLO-II: Decoupling Coverage and {Complexity} with Distinct AP Pair Selection}
{We now discuss the POLO-II algorithm, which increases the localization coverage while keeping {the number of ML cost evaluations constant}.  Similar to POLO-I, it operates in two stages: (i) generating candidate UE positions from differential phase measurements of selected APs; and (ii) obtaining the final UE position by evaluating the ML cost at these candidates, followed by gradient-descent refinement.
We first derive the candidate UE positions for a selected AP quadruplet, and then propose an AP selection strategy to maximize performance.}

\subsection{{Identification of Candidate UE Positions}}
{In POLO-II, the candidate UE positions are estimated using a two-step procedure. In Step 1, for the selected AP quadruplet $\{\text{AP}_{m_1}, \text{AP}_{m_2}, \text{AP}_{m_3}, \text{AP}_{m_4}\}$, the intersection points of the hyperbolas defined by the AP pairs $(\text{AP}_{m_1}, \text{AP}_{m_2})$ and $(\text{AP}_{m_3}, \text{AP}_{m_4})$, are determined. These intersection points serve as the initial candidate set $\mathcal{U}$. However, as shown later, directly using these points for solving \eqref{eq:min_cost} and \eqref{eq:gd} leads to extended high-error regions within the coverage area. To mitigate this, in Step 2, the candidate positions falling in these high-error regions are refined. This is done by using the three differential measurements that can be built from the AP quadruplet taking one AP as reference. Next, we explain each of the steps in detail.}

Using the selected APs, in Step 1 we form two distinct AP pairs: $(\text{AP}_{m_1},\text{AP}_{m_2})$ and $(\text{AP}_{m_3},\text{AP}_{m_4})$. For each pair, the differential phase values
$  \Delta r_{1} = \frac{\lambda}{2\pi}(r_{m_2} - r_{m_1})  $ and $  \Delta r_{2} = \frac{\lambda}{2\pi}(r_{m_4} - r_{m_3}) $ are computed, defining two hyperbolas with respect to the UE position $\mathbf{u}$:
\begin{align}
\mathcal{H}_1 &: \|\mathbf{p}_{m_1} - \mathbf{u}\| - \|\mathbf{p}_{m_2} - \mathbf{u}\| = \Delta r_{1} - \widetilde{z}_{1} \lambda,\label{eq:hyp_3} \\
\mathcal{H}_2 &: \|\mathbf{p}_{m_3} - \mathbf{u}\| - \|\mathbf{p}_{m_4} - \mathbf{u}\| = \Delta r_{2} - \widetilde{z}_{2} \lambda,\label{eq:hyp_4}
\end{align}
where $\widetilde{z}_{1} = z_{m_1} - z_{m_2}$ and $\widetilde{z}_{2} = z_{m_3} - z_{m_4}$, are the differential integer ambiguities.
%
\begin{figure}[t]
	\centering
	\begin{subfigure}[b]{0.8\linewidth}
		\includegraphics[width=0.9\linewidth,height=0.75\linewidth]{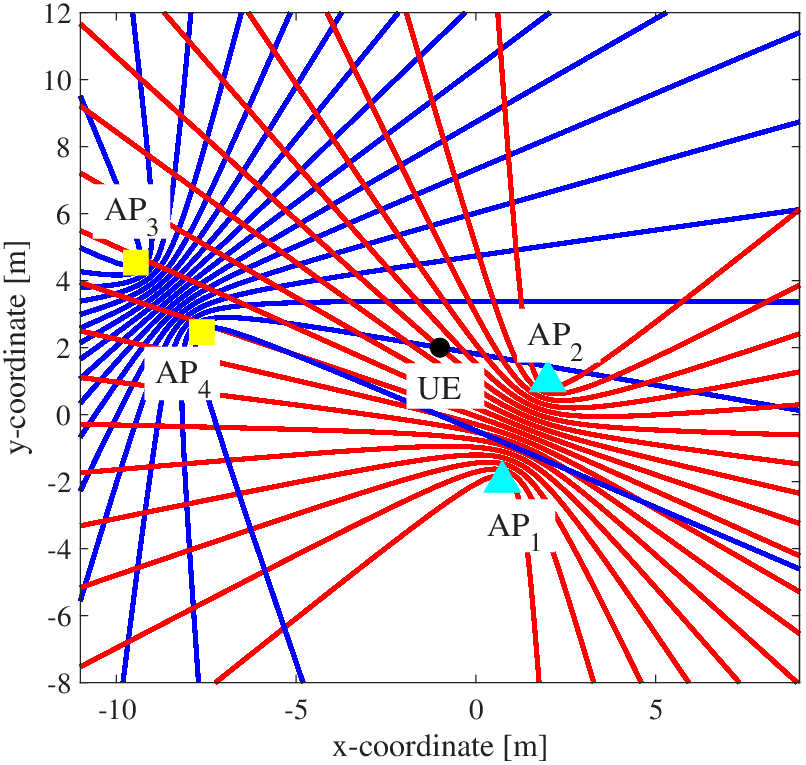}\vspace{0pt}
	\end{subfigure}
	\caption{Illustration of hyperbolas generated from differential phase measurements with two distinct AP pairs in POLO-II. The AP pairs are $\{\text{AP}_1,\text{AP}_2\}$ (green dots) and $\{\text{AP}_3,\text{AP}_4\}$ (yellow dots). The UE location is marked by the black dot. (This color notation is used consistently throughout the paper.)}  
	\label{fig_hyp4}\vspace{0pt}
\end{figure}
%
\begin{figure*}[t]
	\centering
	\begin{subfigure}[b]{0.245\linewidth}
		\includegraphics[width=\linewidth,height=0.8\linewidth]{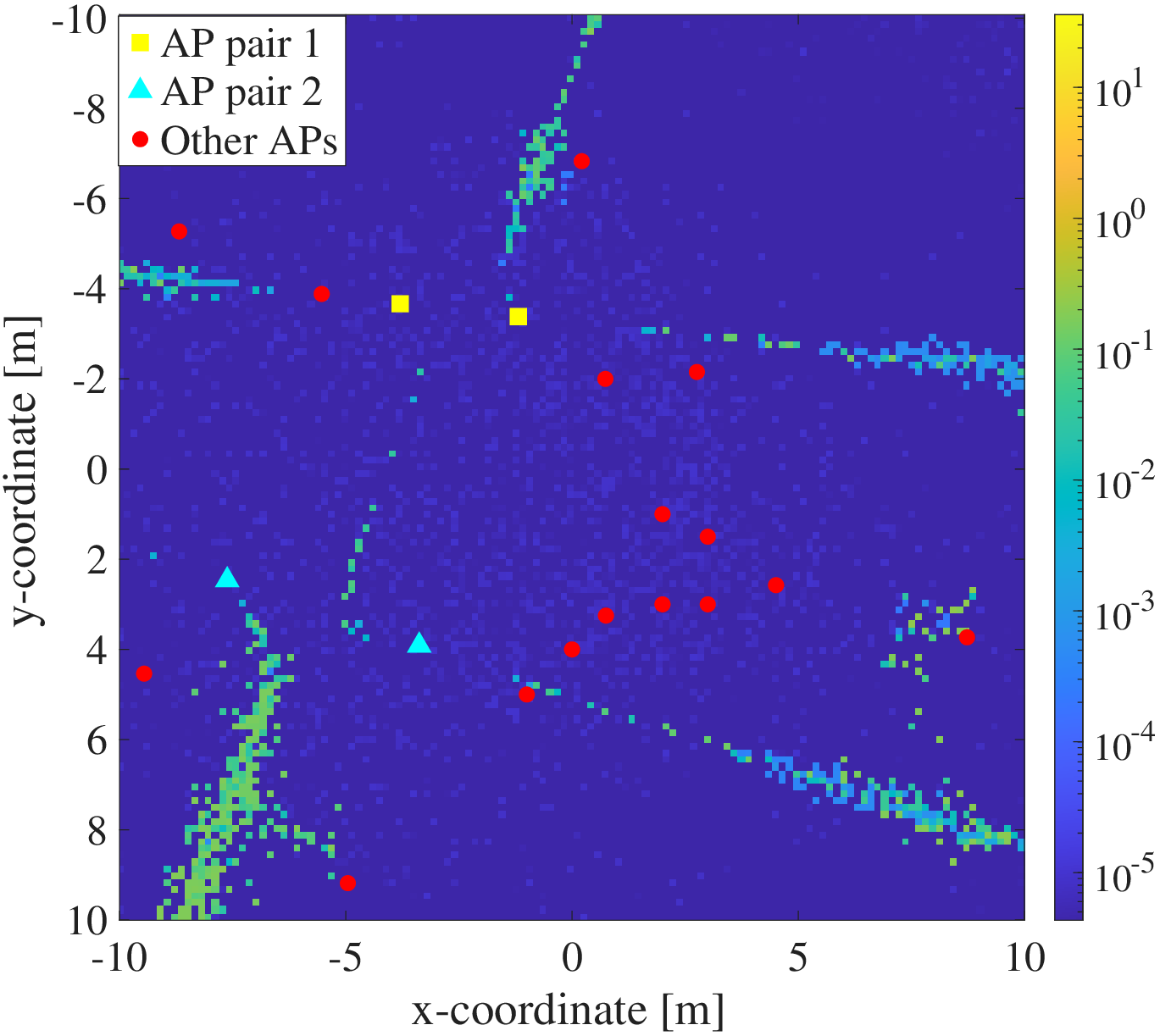}\vspace{0pt}
        \caption{ }  \label{fig_rmse_4AP}
	\end{subfigure}
    \begin{subfigure}[b]{0.245\linewidth}
		\includegraphics[width=\linewidth,height=0.8\linewidth]{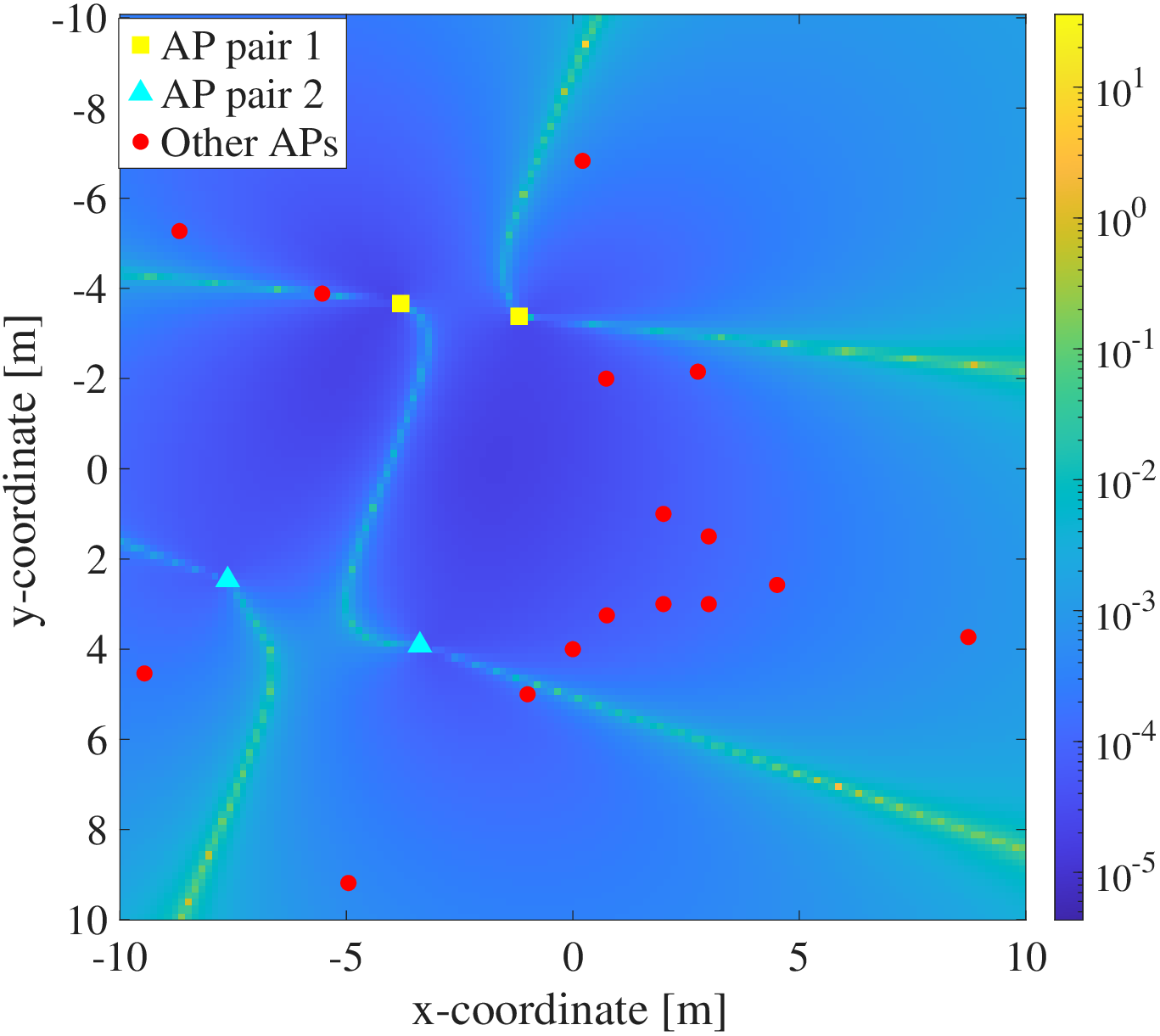}\vspace{0pt}
        \caption{ }  \label{fig_PEB_4AP}
    \end{subfigure}
    \begin{subfigure}[b]{0.245\linewidth}
		\includegraphics[width=\linewidth,height=0.8\linewidth]{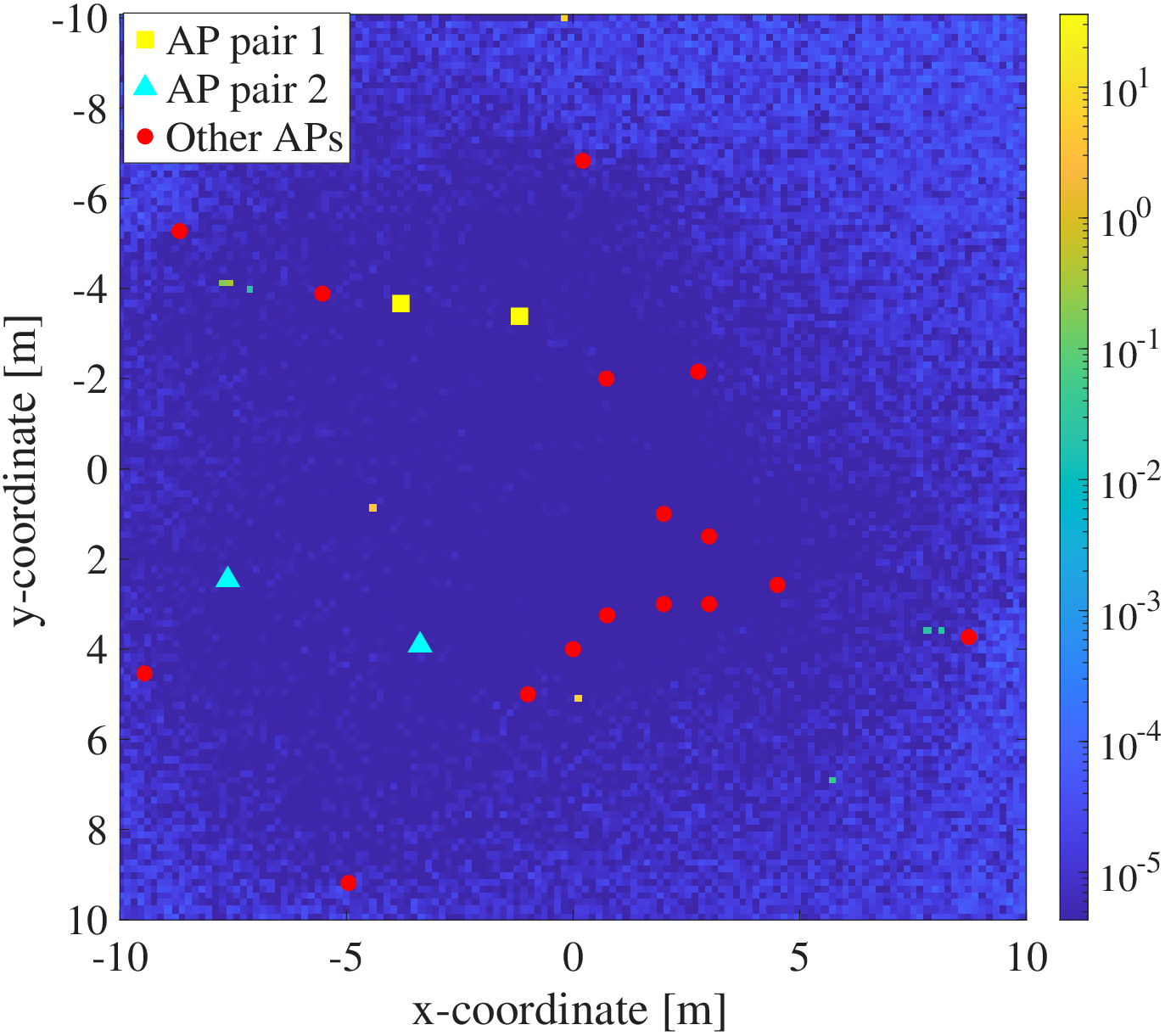}\vspace{0pt}
        \caption{ }  \label{fig_RMSE_4AP_ref}
    \end{subfigure}
       \begin{subfigure}[b]{0.245\linewidth}
		\includegraphics[width=\linewidth,height=0.8\linewidth]{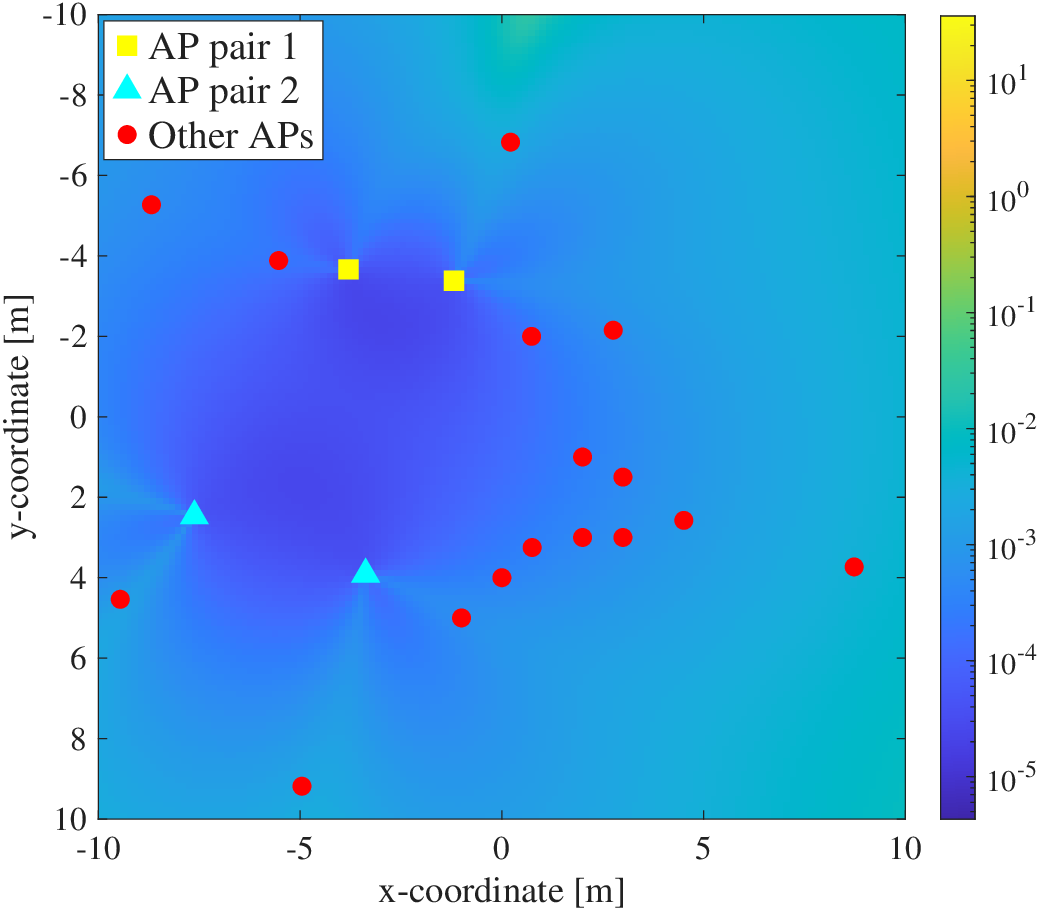}\vspace{0pt}
        \caption{ }  \label{fig_PEB_4AP_all}
    \end{subfigure}
    \vspace{-20pt}
	\caption{ {(a) RMSE map [in m] after Step 1; (b) $\mathrm{PEB}_Q$ map [in m]; (c) RMSE map [in m] after Steps 1 and 2; (d) $\mathrm{PEB}_Q^{\text{ref}}$ map [in m] for different UE locations in the coverage area. The RMSE maps in (a) and (c) are obtained using measurements from all $M$ APs. In (a), the UE candidate set $\mathcal{U}$ is formed from raw hyperbola intersections in Step 1, while in (c) these intersections are refined in Step 2. $\mathrm{PEB}_Q$ in (b) is computed from the EFIM $\mathbf{J}_Q$, representing the sum of pairwise contributions from the selected AP pairs, whereas $\mathrm{PEB}_Q^{\text{ref}}$ in (d) uses EFIM $\mathbf{J}_Q^{\text{ref}}$, which includes the joint information from all four~APs.}\vspace{-20pt}}  
	\label{fig_hmap}
\end{figure*}
In POLO-I, the common reference AP allowed closed-form derivation of hyperbola intersections, simplifying UE candidate computation. In contrast, without a common reference AP, a closed-form solution for $\mathbf{u}$ is unavailable in POLO-II. The  hyperbolas $\mathcal{H}_1$ and $\mathcal{H}_2$ are expressed as a system of two quadratic equations in two variables:
\begin{align}\label{eq:hyp_4apS}
     \!\!\mathcal{H}_1 &:\! A_1 u_x^2 \!+\! B_1 u_x u_y \!+\!C_1 u_y^2 \!+\! D_1 u_x\! +\! E_1 u_y\! +\! F_1 \!=\! 0,\nonumber\\
      \!\!\mathcal{H}_2 &: \!A_2 u_x^2 \!+\! B_2 u_x u_y\! +\!C_2 u_y^2\! + \!D_2 u_x\! +\! E_2 u_y \!+\! F_2 \!=\! 0.\!\!\!\!
\end{align}
The coefficients ${A_i,\cdots,F_i}$ for each hyperbola, $i\in\{1,2\}$, depend on the coordinates of the foci of the hyperbola $(\mathbf{p}_{m_1},\mathbf{p}_{m_2})$ and  $(\mathbf{p}_{m_3},\mathbf{p}_{m_4})$, respectively, and the fixed difference distances $\Delta r_i-\widetilde{z}_i\lambda$. As no closed-form solution for $\mathbf{u}$ exists for this case, we employ iterative numerical methods. {Specifically, we use \texttt{fsolve} function in MATLAB optimization toolbox~\cite{matlab_fsolve}, which employs trust-region and line-search methods to solve nonlinear equations}~{\cite[Sec 11.2]{NUM_OPT}}. Fig.~\ref{fig_hyp4} shows the hyperbolas generated from differential phase measurements of two distinct AP pairs.

Considering selected AP pairs, with intra-AP pair distances $d_{p1}$ and $d_{p2}$, the total number of {hyperbola pairs} can be calculated similar to \eqref{eq:NP1}, and is given~by
\begin{align}\label{eq:NP2}
    N_{\text{P2}} = \frac{4d_{p_1}  d_{p_2}}{\lambda^2}.
\end{align}
Each hyperbola pair yields up to two candidate UE positions. The set $\mathcal{U}$, therefore, contains up to $2N_{\text{P2}}$ possible positions. This completes Step 1 of POLO-II. Step 2 will be detailed in Section \ref{sec:POLOIIrefinement}.

\subsection{PEB for Step 1 and Step 2 of POLO-II}
We note that {the hyperbola intersections derived above can be directly used as the candidate UE set $\mathcal{U}$ to solve \eqref{eq:min_cost} and \eqref{eq:gd}. However, since these intersections use differential phase measurements from only two AP pairs, they do not fully exploit the information from all four APs. Next, we will characterize the PEB of this approach and identify regions of high PEB. Based on this identification, we propose a refinement that fully captures the phase measurements from all four APs.  

\subsubsection{Step 1 PEB Analysis} \label{sec:PEB1POLOII}
To analytically capture the limitation stemming from relying on two differential phase measurement from the two pairs of APs, we next derive the corresponding PEB and EFIM.}
\begin{proposition}\label{prop:EFIM_POLO_II_1}
{Consider a selected AP quadruplet $\mathcal{M} = \{m_i\}_{i=1}^4$, where two distinct AP pairs are formed as $\mathcal{P}_1=\{m_1, m_2\}$ and $\mathcal{P}_2=\{m_3, m_4\}$. The PEB derived from the two differential phase measurements is $\mathrm{PEB}_Q=\sqrt{\mathrm{trace}(\mathbf{J}_Q^{-1})}$, where 
{the EFIM for position is given by}}    
\begin{align}\label{eq:EFIM_POLO_II}
   {\mathbf{J}}_Q(\mathbf{u}) = &B_c \Big(
   \gamma_1 (\mathbf{v}_{m_1} - \mathbf{v}_{m_2})(\mathbf{v}_{m_1} - \mathbf{v}_{m_2})^\top
   + \gamma_2 (\mathbf{v}_{m_3}\nonumber\\
   & - \mathbf{v}_{m_4})(\mathbf{v}_{m_3} - \mathbf{v}_{m_4})^\top
   \Big),
\end{align}
{where $B_c = 8\pi^2 P / (\sigma_n^2 \lambda^2)$, $\gamma_1 = \rho_{m_1}^2 \rho_{m_2}^2 / (\rho_{m_1}^2 + \rho_{m_2}^2)$, and $\gamma_2 = \rho_{m_3}^2 \rho_{m_4}^2 / (\rho_{m_3}^2 + \rho_{m_4}^2)$.}
\end{proposition}
\begin{proof}
{The EFIM $\mathbf{J}_Q$ is derived considering only two independent differential phase measurements, obtained from the AP pairs $\mathcal{P}_1$ and $\mathcal{P}_2$. Consequently, it is the sum of the EFIMs contributed by each AP pair, i.e., $\mathbf{J}_Q = \mathbf{J}_{\mathcal{P}_1} + \mathbf{J}_{\mathcal{P}_2}$, where $\mathbf{J}_{\mathcal{P}_i}$ with $i\in\{1,2\}$ corresponds to the EFIM of the AP pair $\mathcal{P}_i$.}
Each $\mathbf{J}_{\mathcal{P}_i}$ is obtained by substituting the AP set $\mathcal{M}$ with $\mathcal{P}_i$ in \eqref{eq:EFIM_sel_final}.
\end{proof}
{We next present the following corollary to identify the conditions under which $\mathbf{J}_Q$ becomes rank-deficient, corresponding to regions in the coverage with high localization errors.}
\begin{corollary}\label{cor_EFIM_1}
   {The EFIM $\mathbf{J}_Q$ in \eqref{eq:EFIM_POLO_II} becomes rank-deficient when any of the following conditions hold:
   i) $\mathbf{v}_{m_1}=\mathbf{v}_{m_2}$; ii) $\mathbf{v}_{m_3}=\mathbf{v}_{m_4}$;
   iii) ${(\mathbf{v}_{m_1}-\mathbf{v}_{m_2})}/{\norm{(\mathbf{v}_{m_1}-\mathbf{v}_{m_2})}}=\pm{(\mathbf{v}_{m_3}-\mathbf{v}_{m_4})}/{\norm{(\mathbf{v}_{m_3}-\mathbf{v}_{m_4})}}$.}
\end{corollary}
\begin{proof}
   {It follows by substituting the above conditions in \eqref{eq:EFIM_POLO_II} and noticing that $\mathbf{J}_Q$ reduces to a rank-one matrix.}
\end{proof}
{To visualize the regions of high localization errors, we next plot the RMSE of the estimated UE positions in Fig.~\ref{fig_rmse_4AP}. 
The UE position is estimated by solving \eqref{eq:min_cost} and \eqref{eq:gd}, where the candidate set $\mathcal{U}$ is generated by solving \eqref{eq:hyp_3}-\eqref{eq:hyp_4}. 
We also plot $\mathrm{PEB}_Q$ over the coverage area in Fig.~\ref{fig_PEB_4AP}.}
{We observe a strong spatial correlation, similar to POLO-I, between the RMSE and $\mathrm{PEB}_Q$ map. We  notice from Fig.~\ref{fig_hmap} that high localization errors occur along the lines connecting the APs within each pair, extending beyond the segment joining them. This region corresponds to the first and second conditions in Corollary~\ref{cor_EFIM_1}. This pattern is similar to POLO-I. Additionally, high errors occur along the line connecting the two AP pairs, affecting areas both inside and beyond the convex hull formed by the selected APs. This corresponds to the third condition in Corollary~\ref{cor_EFIM_1}. 
Unlike POLO-I, where the area enclosed by the selected AP triplet had low error, the high-error regions now extend into the area enclosed by the selected APs. To address this limitation, the candidate UE positions falling in the high-error regions, identified in Corollary~\ref{cor_EFIM_1}, are refined in the next step.}

\subsubsection{Refinement and Step 2 PEB Analysis} \label{sec:POLOIIrefinement}
Given a set of candidate location $\mathcal{U}$ obtained from solving \eqref{eq:hyp_3}--\eqref{eq:hyp_4}, we define a set $\mathcal{E} \subset \mathcal{U}$ that contains those candidate UE locations that fall in the high-error regions characterized by Corollary~\ref{cor_EFIM_1}. Given a candidate UE location $\widehat{\mathbf{u}}\in{\mathcal{E}}$, 
one AP from the selected quadruplet, say $\text{AP}_{m_1}$, is designated as the reference. The remaining APs are used to form pairs $(\text{AP}_{m_{i+1}}, \text{AP}_{m_1})$ for $i \in \{1,2,3\}$.
Then the corresponding differential phases are computed as $  \Delta \widetilde{r}_{i} = \frac{\lambda}{2\pi}(r_{m_1} - r_{m_{i+1}})  $ which define three hyperbolas ($i\in\{1,2,3\}$):
\begin{align}
\mathcal{H}_{i} &: \|\mathbf{p}_{m_{i+1}} - \mathbf{u}\| - \|\mathbf{p}_{m_1} - \mathbf{u}\| = \Delta \widetilde{r}_{i} - \widehat{z}_{i}(\widehat{\mathbf{u}}) \lambda,  \label{eq:hyp_5} 
\end{align}
where $\widehat{z}_{i}(\widehat{\mathbf{u}}) = z_{m_{i+1}}(\widehat{\mathbf{u}})-z_{m_1}(\widehat{\mathbf{u}})$ defines the differential integer ambiguity at the candidate UE location $\widehat{\mathbf{u}}\in{\mathcal{E}}$. {The differential ambiguity $\widehat{z}_{i}(\widehat{\mathbf{u}})$  for $i=1$ is obtained using \eqref{eq:hyp_3}, while for $i=2$ and $i=3$ is obtained by substituting $\mathbf{p}_{m_2}$ in \eqref{eq:hyp_3} with $\mathbf{p}_{m_3}$ and $\mathbf{p}_{m_4}$, respectively.}
These hyperbolas have a similar structure to \eqref{eq:hyp_1}–\eqref{eq:hyp_2}, and their intersection points can be approximated in closed form. The derivation follows the approach in Appendix~\ref{app_hyp_intersection} and is omitted for brevity. {Since three hyperbolas are involved in \eqref{eq:hyp_5}, their intersections may not produce an exact solution, but rather a least-squares approximation.} 
This solution  provides refined estimates of the candidate UE locations $\widehat{\mathbf{u}}$, based on the full three differential phase measurements, as opposed to two differential measurements considered in Section \ref{sec:PEB1POLOII}. 
After refining the candidate UE locations in the high-error regions, we obtain an updated set of candidate UE positions $\widehat{\mathcal{U}}$ (we call this Step 2 of POLO-II). This set is used in \eqref{eq:min_cost} and \eqref{eq:gd} to estimate the final UE position. The corresponding RMSE map is shown in Fig.~\ref{fig_RMSE_4AP_ref}. Compared to the map using the raw candidate set $\mathcal{U}$ after Step~1 (Fig.~\ref{fig_rmse_4AP}), the high-error regions nearly disappear after refinement, yielding almost complete localization coverage.


The PEB derived in Proposition \ref{prop:EFIM_POLO_II_1} is no longer a valid metric to analyze the RMSE in Fig.~\ref{fig_RMSE_4AP_ref}, as one additional phase measurement is used. To verify from a PEB perspective that the three cases identified in Corollary \ref{cor_EFIM_1} are no longer present, we have the following proposition. 
\begin{proposition}\label{prop:EFIM_POLO_II_refined}
Consider the selected AP quadruplet $\mathcal{M} = \{m_i\}_{i=1}^4$, and assume that the UE lies in the high-error regions characterized in Corollary~\ref{cor_EFIM_1}. 
Then, the corresponding EFIM ${\mathbf{J}}_{Q}^{\text{ref}}(\mathbf{u})$ for position is expressed as follows:
    i) when $\mathbf{v}_{m_1} = \mathbf{v}_{m_2} = \mathbf{v}$, the EFIM is given by
    \begin{align}\label{eq:EFIM_sp_case_POLO_II}
        {\mathbf{J}}_{Q}^{\text{ref}}(\mathbf{u})
        &= K_c \Big( 
        \sum_{i=3}^{4} 
       \rho_i^2(\rho_1^2 + \rho_2^2) (\mathbf{v} - \mathbf{v}_{m_i})(\mathbf{v} - \mathbf{v}_{m_i})^\top
        \nonumber\\[-0.3em]
        &\quad
        + \rho_3^2 \rho_4^2 (\mathbf{v}_{m_3} - \mathbf{v}_{m_4})(\mathbf{v}_{m_3} - \mathbf{v}_{m_4})^\top 
        \Big);
    \end{align}
     ii) the case $\mathbf{v}_{m_3}=\mathbf{v}_{m_4}$ results in an equivalent expression;
    iii) when $\frac{\mathbf{v}_{m_1} - \mathbf{v}_{m_2}}
    {\|\mathbf{v}_{m_1} - \mathbf{v}_{m_2}\|}
    = 
    \pm
    \frac{\mathbf{v}_{m_3} - \mathbf{v}_{m_4}}
    {\|\mathbf{v}_{m_3} - \mathbf{v}_{m_4}\|},$ the EFIM is given by
    \begin{align}\label{eq:EFIM_sp_case_POLO_II_2}
        {\mathbf{J}}_Q^{\text{ref}}(\mathbf{u})
        &= K_c \Big(
        (\rho_1^2 \rho_2^2 + \rho_3^2 \rho_4^2)
        (\mathbf{v}_{m_1} - \mathbf{v}_{m_2})
        (\mathbf{v}_{m_1} - \mathbf{v}_{m_2})^\top
        \nonumber\\[-0.3em]
        &
        \!\!\!\!\!\!+ \sum_{i=1}^{2}\rho_i^2 
        \sum_{j=3}^{4}\rho_j^2 
        (\mathbf{v}_{m_i} - \mathbf{v}_{m_j})
        (\mathbf{v}_{m_i} - \mathbf{v}_{m_j})^\top
        \Big). \!\!\!\!\!\!
    \end{align}
Hence, in all 3 cases from Corollary~\ref{cor_EFIM_1}, the EFIM is no longer rank-deficient, and therefore the corresponding PEB $\mathrm{PEB}_Q^{\text{ref}}=\sqrt{\mathrm{trace}((\mathbf{J}_Q^{\text{ref}})^{-1})}$ is bounded. 
\end{proposition}
\begin{proof}
   {Eq. \eqref{eq:EFIM_sp_case_POLO_II} and \eqref{eq:EFIM_sp_case_POLO_II_2} can be obtained by substituting $\mathbf{v}_{m_1} = \mathbf{v}_{m_2} = \mathbf{v}$ and $\frac{\mathbf{v}_{m_1} - \mathbf{v}_{m_2}}{\|\mathbf{v}_{m_1} - \mathbf{v}_{m_2}\|}= \pm\frac{\mathbf{v}_{m_3} - \mathbf{v}_{m_4}}
    {\|\mathbf{v}_{m_3} - \mathbf{v}_{m_4}\|}$ respectively, in the expression of $\mathbf{J}_Q^{\text{ref}}$, which can be obtained by extending Proposition~\ref{prop_EFIM} for $\mathcal{M}=\{m_i\}_{i=1}^4$.}
\end{proof}

Note that the only degenerate case that arises  is when the UE lies at the intersection of the lines connecting any two of the four selected APs, as formalized next. 
%
\begin{corollary}
When the UE is located at the intersection of the two lines formed by the AP pairs $(m_1,m_2)$ and $(m_3,m_4)$, i.e., $\mathbf{v}_{m_1}=\mathbf{v}_{m_2}=\mathbf{v}$ and $\mathbf{v}_{m_3}=\mathbf{v}_{m_4}=\widetilde{\mathbf{v}}$, the corresponding EFIM simplifies to
\begin{align}
{\mathbf{J}}_Q^{\text{ref}}(\mathbf{u}) = \widetilde{K}_c(\mathbf{v}-\widetilde{\mathbf{v}})(\mathbf{v}-\widetilde{\mathbf{v}})^\top,
\end{align}
where $\widetilde{K}_c = K_c(\widetilde{\beta}_3 + \widetilde{\beta}_4)$. 
In this case, the EFIM is rank-deficient, resulting in large localization error.
\end{corollary}
\begin{proof}
The result follows directly by substituting $\mathbf{v}_{m_3}=\mathbf{v}_{m_4}=\widetilde{\mathbf{v}}$ in \eqref{eq:EFIM_sp_case_POLO_II}.
\end{proof}

A key distinction between POLO-I and POLO-II lies in the rank deficiency of their EFIMs. In POLO-I, the selected triplet EFIM $\mathbf{J}_T(\mathbf{u})$ from \eqref{eq:EFIM_sp_case_POLO_I} is rank-deficient along the lines connecting any pair of selected APs, except the segment between them, leading to extended regions of unbounded $\mathrm{PEB}_T$ (see Fig.~\ref{fig_sample}). In contrast, in POLO-II, the EFIM ${\mathbf{J}}_Q^{\text{ref}}(\mathbf{u})$ is rank-deficient only at the isolated intersection points of these lines, specifically three such points per AP quadruplet, resulting in negligible high-error regions. This behavior is illustrated in Fig.~\ref{fig_PEB_4AP_all}, which shows the corresponding ${\mathrm{PEB}}_Q^{\text{ref}}=\sqrt{(\mathbf{J}_Q^{\text{ref}})^{-1}}$. 
{The RMSE after  refinement in Step~2, as illustrated in Fig.~\ref{fig_RMSE_4AP_ref},  achieves low values across most of the coverage area, analogous to the behavior of $\mathrm{PEB}_Q^{\text{ref}}$.}


\begin{remark}
 In principle, Step 2 can be applied directly, by skipping Step 1, to obtain the set of candidate UE locations. However, as in POLO-I, the number of hyperbola intersections increases with the distance between the reference AP and the other three APs. To keep this number feasible, all three APs would need to be located close to the reference AP, which is difficult in practice. In contrast, identifying two closely spaced AP pairs, as done in Step 1, is more practical. Step 1 produces a smaller candidate set, which is then~refined~in~Step~2.
\end{remark}

\subsection{{AP Selection Strategy for POLO-II}}\label{sec_cov_comp_POLO_II}
{We observe from Fig.~\ref{fig_PEB_4AP_all} that $\mathrm{PEB}^{\text{ref}}_Q$ attains lower values within the convex hull formed by the selected APs, while outside this region it increases but remains bounded. Extending this low-$\mathrm{PEB}^{\text{ref}}_Q$ region improves the reliability of UE position estimation over a wider area.}
In POLO-I, extending the low-$\mathrm{PEB}_T$ region required increasing the intra-AP pair distance, which increased the number of ML cost evaluations. 
In contrast, POLO-II uses two distinct AP pairs in Step~1 to obtain the initial set of UE candidate positions. The low-$\mathrm{PEB}_Q$ region here can be expanded either by increasing the intra-AP pair distance (between APs in a pair) or the inter-AP pair distance (between the midpoints of the two AP pairs). While larger intra-AP pair distances introduce additional hyperbolas, increasing only the inter-AP pair distance enhances the coverage area without raising the number of intersections, effectively \textit{decoupling coverage from the number of ML cost evaluations}. However, since the hyperbola intersections are computed numerically, the reduction in number of cost evaluations does not always imply a proportional complexity reduction. 

Based on these insights, we propose an AP selection strategy for POLO-II in Algorithm \ref{alg:ap_selection_polo_II}. This strategy balances performance and complexity by first eliminating AP quadruplets that are computationally expensive, i.e. with larger intra-pair distances, and then selecting those that enable higher localization coverage area with larger inter-AP pair distances.

\begin{algorithm}[t]
\caption{{POLO-II: AP Selection Strategy }}
\label{alg:ap_selection_polo_II}
\begin{algorithmic}[1]
\STATE \textbf{Input:} Set of all AP positions $\{\mathbf{p}_i\}_{i=1}^M$, and $\gamma=$ AP distance threshold.
\STATE \textbf{Output:} Selected AP quadruplet $\{(\widehat{i}_1, \widehat{i}_2), (\widehat{i}_3, \widehat{i}_4)\}$
\vspace{0.5em}
\STATE Define set of all AP quadruplets with disjoint AP pairs $\mathcal{Q}=\{(i_1,i_2),(i_3,i_4)|i_m\in\{1,\cdots,M\}, i_m\neq i_n,\forall m\neq n\}$
\FOR {each $\{(i_1, i_2), (i_3, i_4)\} \in \mathcal{Q}$} 
\STATE Compute: $d_{\text{intra}}^{((i_1, i_2), (i_3, i_4))} = \frac{ \|\mathbf{p}_{i_1} - \mathbf{p}_{i_2}\| + \|\mathbf{p}_{i_3} - \mathbf{p}_{i_4}\| }{2} $
\IF{$d_{\text{intra}}^{((i_1,i_2),(i_3,i_4))}<\gamma$}
        \STATE Store the quadruplet  $\{(i_1,i_2),(i_3,i_4)\}\in \mathcal{Q}_\text{valid}$
        \STATE Compute $d_{\text{inter}}^{((i_1, i_2), (i_3, i_4))} = \left\| \frac{\mathbf{p}_{i_1} + \mathbf{p}_{i_2}}{2} - \frac{\mathbf{p}_{i_3} + \mathbf{p}_{i_4}}{2} \right\|$
    
    \ENDIF
\ENDFOR

\STATE \textbf{Return} $\{(\widehat{i}_1,\widehat{i}_2),(\widehat{i}_3,\widehat{i}_4)\}=\!\!\!\!\!\!\underset{(i_1, i_2), (i_3, i_4)\in\mathcal{Q}_{\text{valid}}}{\arg\max}\!\!\!d_{\text{inter}}^{((i_1, i_2), (i_3, i_4))}$
\end{algorithmic}
\end{algorithm}

\section{Complexity Analysis}
This section analyzes the computational complexity of the proposed POLO-I and POLO-II algorithms. As a baseline, we first briefly discuss the complexity of conventional UE localization using MLE with EGS~\cite{RS_UPLINK_JOURNAL,DOWNLINK_DIST,GRID_SEARCH_NAS,MLE_EGS}.

\subsection{MLE using EGS}
The EGS approach first evaluates the ML cost function over a dense grid covering the coverage area~\cite{GRID_SEARCH_NAS,MLE_EGS}, then refines the best grid point using gradient descent. Its overall computational complexity is~\cite{GRID_SEARCH_NAS,MLE_EGS}
\begin{align}
C_{\text{EGS}} = C_{\text{ML}} N_{\text{grid}} + C_{\text{GD}},
\end{align}
where $N_{\text{grid}} = A/(k^2\lambda^2)$ is the number of grid points for coverage area $A$ and grid resolution $k\lambda \times k\lambda$ ($0<k\leq 1$),  $C_{\text{ML}}$ is the complexity of evaluating the ML function at one grid point, and $C_{\text{GD}}$ is the complexity of applying gradient descent.
At each grid point $\mathbf{u}$, evaluating the ML cost involves computing $\widehat{\phi}(\mathbf{u})$ in \eqref{eq:delta_phi} and $C(\mathbf{u})$ in \eqref{eq:cost_final}, with overall complexity $C_{\text{ML}} = \mathcal{O}(M)$.
Gradient descent, assuming convergence in $L$ iterations, has computational complexity of $C_{\text{GD}} = \mathcal{O}(L M)$. The overall complexity, therefore, becomes 
$C_{\text{EGS}} = \mathcal{O} ( M ( {A}/{(k^2\lambda^2)} + L ) )$.

\subsection{POLO-I}
We now analyze the complexity of the POLO-I algorithm, which comprises the following three steps:
\begin{itemize}
    \item {Step 1:} Compute the intersection points of the hyperbolas $\mathcal{H}_1$ and $\mathcal{H}_2$, defined in \eqref{eq:hyp_1} and \eqref{eq:hyp_2}, respectively. The computational cost of this step is denoted as $C_{\text{IP1}}$.
    \item {Step 2:} Evaluate the ML cost function in \eqref{eq:cost_final} at each candidate position. The complexity of evaluating the ML cost, as discussed previously, is $C_{\text{ML}}=\mathcal{O}(M)$.
    \item {Step 3:} Apply gradient descent for final refinement, whose complexity with $L$ iterations is $C_{\text{GD}}=\mathcal{O}(LM)$.
\end{itemize}
The computational complexity of POLO-I is thus given as
\begin{align}\label{eq:comp_POLOI}
    C_{\text{POLO-I}} = (C_{\text{IP1}} + 2C_{\text{ML}}) N_{\text{P1}} + C_{\text{GD}},
\end{align}
where $N_{\text{P1}}$ is the number of {possible hyperbola pairs. The factor of $2$ is included with $C_{\text{ML}}$ as each hyperbola pair, as previously discussed, yields up to two potential candidate UE positions, at which the ML cost must be evaluated.} 
The number of hyperbola pairs $N_{\text{P1}}$ is given in \eqref{eq:NP1}. 
Since closed form solutions for the hyperbola intersection points are available, the complexity of this step is $C_{\text{IP1}} = \mathcal{O}(1)$. 
The overall computational complexity of the POLO-I algorithm is thus
    $C_{\text{POLO-I}} = \mathcal{O}( M({8d_{s_1}d_{s_2}}/{\lambda^2} + L) )$.

\subsection{POLO-II}
The computational complexity for POLO-II, on the lines similar to POLO-I, is expressed as: 
\begin{align}\label{eq:comp_POLOII}
    C_{\text{POLO-II}} = (C_{\text{IP2}} + 2C_{\text{ML}})  N_{\text{P2}} + C_{\text{GD}},
\end{align}
{where $N_{\text{P2}}$, expressed in \eqref{eq:NP2}, is the number of hyperbola pairs, and $C_{\text{IP2}}$ is the complexity of identifying the intersection of two hyperbolas in POLO-II. 
The key distinction between POLO-I and POLO-II lies in how candidate UE locations are computed, while the complexities $C_{\text{ML}}$ and $C_{\text{GD}}$ remain unchanged.} 
In POLO-II, the UE candidate locations are computed in two steps. Step 1 numerically solves \eqref{eq:hyp_3}-\eqref{eq:hyp_4} for $\mathbf{u}$, with computational cost scaling with the number of iterations.
Step 2 refines a subset of these UE candidates using \eqref{eq:hyp_5}, which admits closed form solution, and hence adds negligible complexity.
Assuming $L_\text{IP2}$ iterations in Step 1, the overall complexity of computing UE candidate positions in POLO-II is $C_{\text{IP2}}=\mathcal{O}(L_\text{IP2})$~\cite{NUM_OPT}. In practice, $L_\text{IP2}$ ranges between $20$ and $30$.
The total computational complexity of POLO-II is thus, $    C_{\text{POLO-II}} = \mathcal{O}( M({8d_{p_1}d_{p_2}}/{\lambda^2}+ L) + L_\text{IP2}{4d_{p_1}d_{p_2}}/{\lambda^2})$.

\begin{table}[t]
\centering
\small
\renewcommand{\arraystretch}{1.5}
\begin{tabular}{|c|c|c|c|}
\hline
{\textbf{Metric}} & {\textbf{EGS}} & {\textbf{POLO-I}} & {\textbf{POLO-II}} \\
\hline
\makecell{Complexity of \\ computing hyperbola \\ intersections} 
& - 
& $\mathcal{O}(1)$ 
& $\mathcal{O}(L_\text{IP2})$ \\
\hline
\makecell{No. of ML \\ evaluations} 
& $\displaystyle \frac{A}{k^2\lambda^2}$ 
& $\displaystyle \frac{8d_{s_1}d_{s_2}}{\lambda^2}$ 
& $\displaystyle \frac{8d_{p_1}d_{p_2}}{\lambda^2}$ \\
\hline
\makecell{ML cost evaluation \\ complexity} 
& $\mathcal{O}(M)$ 
& $\mathcal{O}(M)$ 
& $\mathcal{O}(M)$ \\
\hline
\makecell{Gradient descent \\ complexity} 
& $\mathcal{O}(LM)$ 
& $\mathcal{O}(LM)$ 
& $\mathcal{O}(LM)$ \\
\hline
\end{tabular}
\caption{Comparison of complexity across localization algorithms.
{ Here, $A$ is coverage area; $d_{s_i}$ and $d_{p_i}$ are intra-pair distances of $i$-th pair, and $N_{\text{P1}}$ and $N_{\text{P2}}$ are numbers of hyperbola pairs in POLO-I and POLO-II, respectively; $N_{\text{grid}}$ is number of ML cost evaluations in EGS; $L$ is  number of gradient-descent iterations; and $L_{\text{IP2}}$ is number of iterations to compute hyperbola intersections~in POLO-II.}}
\label{tab:complexity_comparison}
\end{table}

\subsubsection{Comparison of Computational Complexities}
{Key complexity metrics for each localization algorithm is summarized in Table~\ref{tab:complexity_comparison}. We see that the EGS method is computationally expensive due to large number of grid evaluations. In contrast, POLO-I and POLO-II has reduced complexity, as their evaluations scale with intra-AP pair distances ($d_{s_1}d_{s_2}$ or $d_{p_1}d_{p_2}$), rather than the total coverage area $A$, as in the EGS method. }

{POLO-I achieves lower computational complexity by providing closed-form hyperbola intersections, whereas POLO-II incurs a slightly higher overhead due to their numerical computation. However, in POLO-I, expanding coverage requires increasing intra-AP pair distances, which raises the number of ML cost evaluations, and hence the complexity. In contrast, POLO-II can enhance coverage by increasing inter-AP pair distances, which keeps the number of ML cost evaluations constant. Consequently, POLO-I offers ultra-low complexity with limited coverage, while POLO-II achieves broader coverage at moderate computational cost.}

\vspace{-7pt}
\section{Numerical Results}\vspace{-3pt}\label{sec_sim}
In this section, we present numerical simulations to evaluate and compare the localization accuracy and computational complexity of the proposed POLO-I and POLO-II algorithms, and benchmark them against the conventional EGS method.

\vspace{-7pt}
\subsection{Scenario}\label{sec_sim_scenario}
We consider a network of $M = 20$ single-antenna APs randomly deployed within a $20$ m$\times 20$ m coverage area. Unless specified, the UE transmits a unit pilot symbol $s$ with transmit power $P = 5$ dBm and experiences a phase offset of $\phi = 10^\circ$ relative to the AP network. The system operates at a carrier frequency of $f_0 = 3.5$ GHz. The transmit and receive antenna gains $G_{\text{tx}}$ and $G_{\text{rx}}$ are set to $1$. The noise power is calculated over a reference bandwidth of $W = 120$ kHz, corresponding to a single subcarrier, and is given by $\sigma_n^2 = k_B T_{\rm{tn}} W$, where $k_B = 1.38 \times 10^{-23}$ J/K is the Boltzmann constant, and $T_{\rm{tn}} = 290$ K is the thermal noise temperature.

\vspace{-7pt}
\subsection{Localization Accuracy Analysis}

\subsubsection{RMSE Evaluation of Proposed Algorithms}
We evaluate the performance of the proposed POLO-I and POLO-II algorithms by comparing the RMSE of the estimated UE position with the PEB in Fig.~\ref{fig_rmse_peb_tx_pwer}, as a function of the UE transmit power. 
{This PEB is evaluated considering measurements from all $M$ APs, and is denoted as $\mathrm{PEB}_F$.} 
{For this study, we employ the AP selection strategy in Algorithm \ref{alg:ap_selection} for POLO-I and Algorithm \ref{alg:ap_selection_polo_II} for POLO-II, and consider the UE to be located at $[1,2]^\top$.}
For benchmarking, we include the RMSE of the MLE obtained using the EGS method~{\cite{GRID_SEARCH_NAS,MLE_EGS}}. 
To implement EGS, the coverage area is divided into a grid with resolution $0.1\lambda \times 0.1\lambda$, and the ML cost function is evaluated at each grid point.
We notice that all three approaches, POLO-I, POLO-II, and EGS, achieve the PEB at sufficiently high transmit powers, confirming the asymptotic efficiency of the proposed approaches. We notice that, among them, EGS achieves the PEB at the lowest transmit power, followed by POLO-II. POLO-I requires the highest transmit power to reach the bound. 
{Although this observation corresponds to a specific UE location, the same trend holds for most UE locations in the coverage area. A comprehensive evaluation of their performance across the coverage area and under different AP selections is presented in the following~sections.}

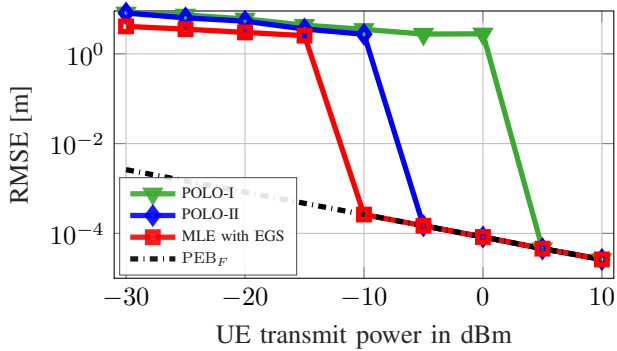
\begin{figure}[t]
    \centering
%
%
\definecolor{mycolor1}{rgb}{0.23137,0.66667,0.19608}%
\definecolor{mycolor2}{rgb}{0.12941,0.12941,0.12941}%
\begin{tikzpicture}

\begin{axis}[%
width=0.75\columnwidth,
height=1.4in,
at={(0.905in,0.583in)},
scale only axis,
xmin=-31,
xmax=11,
xlabel style={font=\color{white!15!black}},
xlabel={UE transmit power in dBm},
ymode=log,
ymin=1e-05,
ymax=9,
yminorticks=true,
ylabel style={font=\color{white!15!black}},
ylabel={RMSE  [m]},
axis background/.style={fill=white},
xmajorgrids,
ymajorgrids,
yminorgrids,
legend style={font=\footnotesize,at={(0.01,0.01)},nodes={scale=0.75, transform shape}, anchor=south west, legend cell align=left, align=left, draw=white!25!black, opacity=0.9}
]
\addplot [color=mycolor1, line width=2.0pt, mark size=2.7pt, mark=triangle, mark options={solid, rotate=180, mycolor1}]
  table[row sep=crcr]{%
-30	8.14575\\
-25	7.3423\\
-20	6\\
-15	4.3542\\
-10	3.51081462342827\\
-5	2.7546\\
0	2.78871876932852\\
5	4.57094028025e-05\\
10	2.62810587894e-05\\
};
\addlegendentry{POLO-I}

\addplot [color=blue, line width=2.0pt, mark size=2.7pt, mark=diamond, mark options={solid, blue}]
  table[row sep=crcr]{%
-30	8.10588084814312\\
-25	6.324\\
-20	5.33410620665458\\
-15	3.5643\\
-10	2.73203496508654\\
-5	0.000147789254270107\\
0	8.3108005095e-05\\
5	4.57094028025e-05\\
10	2.62810587894e-05\\
};
\addlegendentry{POLO-II}

\addplot [color=red, line width=2.0pt, mark size=2pt, mark=square, mark options={solid, red}]
  table[row sep=crcr]{%
-30	4.09575398799835\\
-25	3.5431\\
-20	3\\
-15	2.54312\\
-10	0.000262810587894\\
-5	0.000147789254270107\\
0	8.3108005095e-05\\
5	4.57094028025e-05\\
10	2.62810587894e-05\\
};
\addlegendentry{MLE with EGS}

\addplot [color=black, dashdotted, line width=2.0pt]
  table[row sep=crcr]{%
-30	0.002628105878943\\
-25	0.00147789254270548\\
-20	0.000831080050954\\
-15	0.00046735065719243\\
-10	0.000262810587894\\
-5	0.000147789254270107\\
0	8.3108005095e-05\\
5	4.57094028025e-05\\
10	2.62810587894e-05\\
};
\addlegendentry{$\mathrm{PEB}_F$}

\end{axis}

\end{tikzpicture}%
    \caption{RMSE of the UE position estimate using POLO-I, POLO-II and EGS, and  $\mathrm{PEB}_F$.  $\mathrm{PEB}_F$ represents the PEB evaluated using the measurements from all $M$ APs. All methods achieve  $\mathrm{PEB}_F$ at high transmit powers. {This study is conducted when the UE is located at $[1,2]^\top$, however same trend holds for most of UE locations within the coverage area.}  }
    \label{fig_rmse_peb_tx_pwer}
\end{figure}

\begin{figure}[t]
\vspace{-10pt}
	\centering
	\begin{subfigure}[b]{\linewidth}\centering
		\includegraphics[width=0.8\linewidth,height=0.7\linewidth]{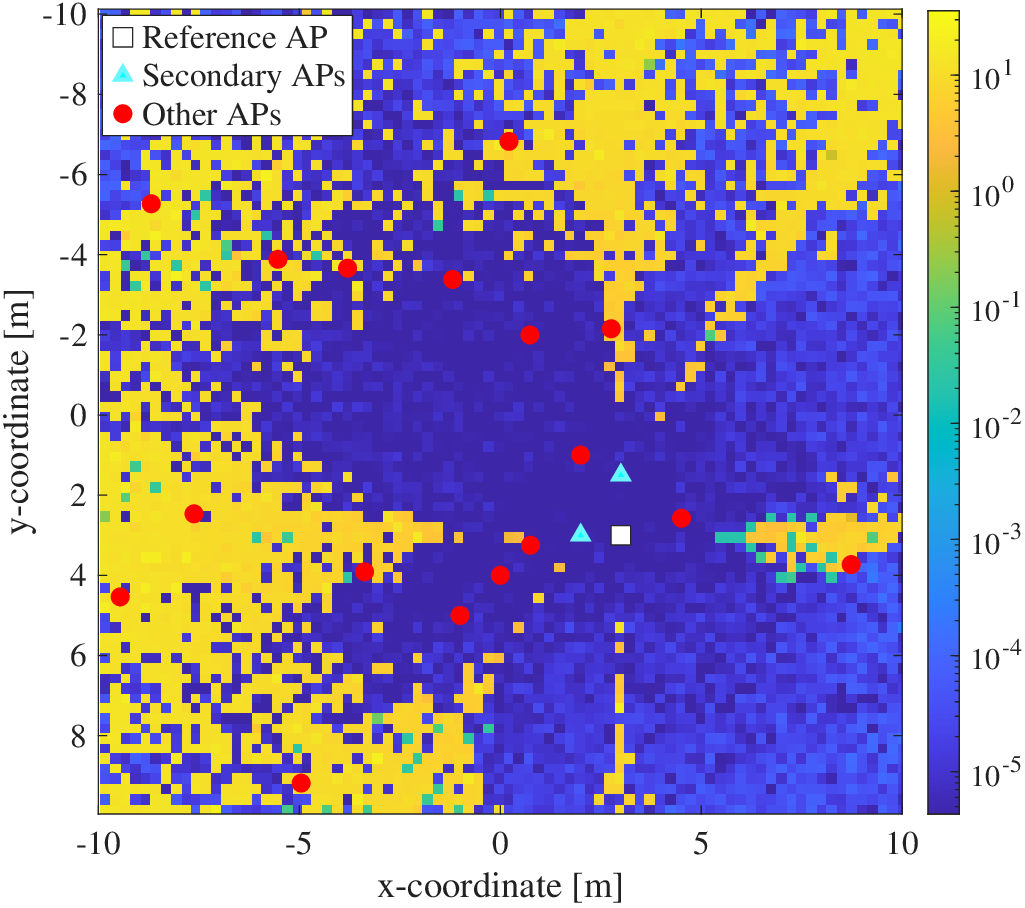}\vspace{0pt}
        \caption{ }  \label{fig_rmse_closest}
	\end{subfigure}
    	\begin{subfigure}[b]{\linewidth}\centering
		\includegraphics[width=0.8\linewidth,height=0.7\linewidth]{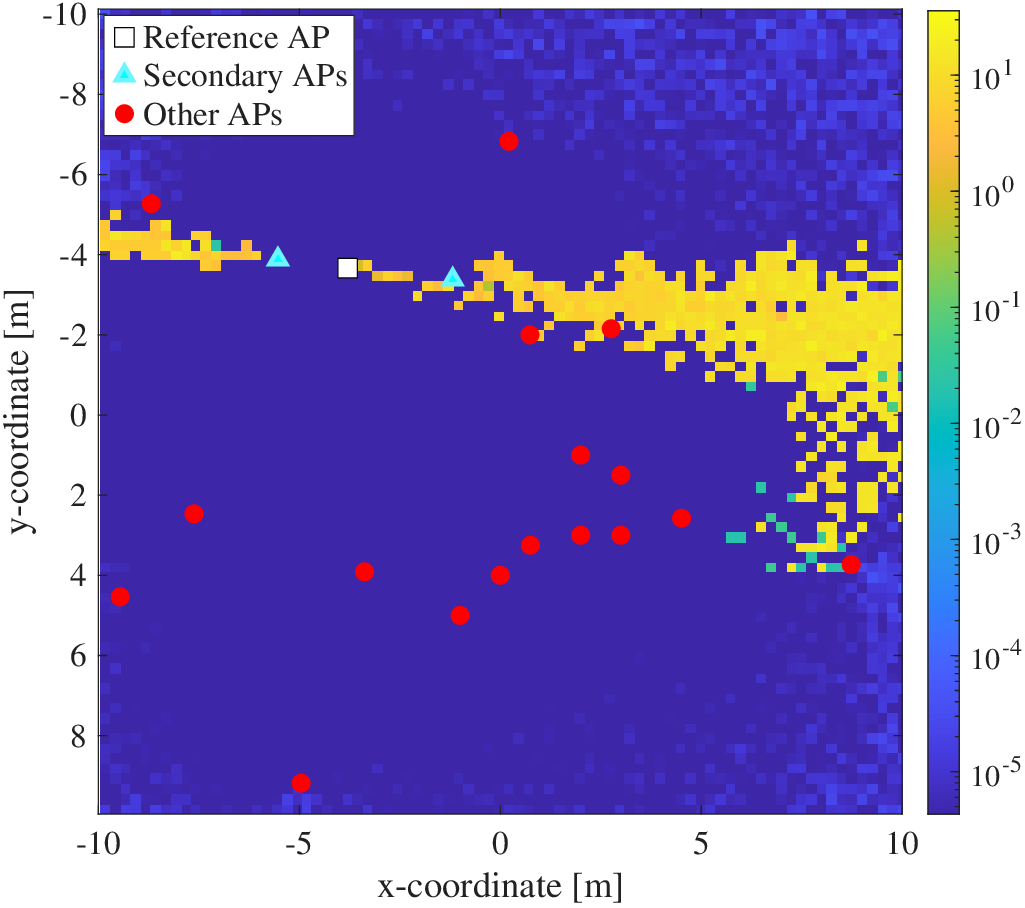}\vspace{0pt}
        \caption{ }  \label{fig_rmse_col}
	\end{subfigure}
	\caption{RMSE [in m] maps for POLO-I using: (a) Strategy 1: choosing the three closest APs; (b) Strategy 2: PEB-induced AP selection. Strategy 1 leads to larger high-error (yellow) regions, while strategy 2 significantly reduces high-error areas, achieving performance closer to  $\mathrm{PEB}_F$. }  
	\label{fig_sel}\vspace{0pt}
\end{figure}
\subsubsection{Impact of AP Selection Strategies in POLO-I}
We illustrate the RMSE heat maps in Fig.~\ref{fig_sel} for AP selection strategies 1 and 2 using the POLO-I algorithm. These maps are generated by varying the UE position across the coverage area and computing the RMSE at each location. Two distinct regions are observed in the heat maps: i) yellow, indicating high localization errors, and ii) blue, indicating low RMSE approaching the PEB. Strategy 1, which selects the three closest APs to minimize complexity, exhibits a larger high-error (yellow) region than Strategy 2, which selects APs based on their geometric configuration. This shows that Strategy 2 provides a larger area of high-accuracy localization at a given transmit power, demonstrating its superiority in coverage.
\begin{figure}[t]
\vspace{-45pt}
    \centering
    \input{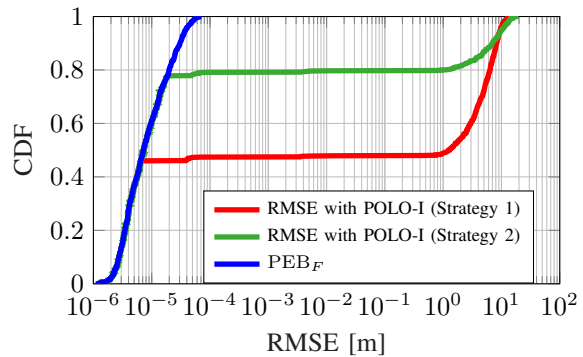}
    \caption{CDF of the RMSE values corresponding to the maps in Fig. \ref{fig_sel}. Strategy 2 achieves RMSE close to the PEB in about $80\%$ of the area, compared to only $45\%$ for Strategy 1.}
    \label{fig_rmse_cdf_POLO}
\end{figure}

{To further illustrate the difference, Fig.~\ref{fig_rmse_cdf_POLO} shows the CDF of RMSE values from the heat maps of both strategies, alongside the CDF of the $\mathrm{PEB}_F$ evaluated at each UE location. With Strategy 1, only $\sim 45\%$ of the area achieves RMSE close to  $\mathrm{PEB}_F$, while Strategy 2 reaches  $\mathrm{PEB}_F$ in $\sim 80\%$ of the region, for a UE transmit power of $P = 5$~dBm. Increasing transmit power raises the fraction of the area where RMSE approaches  $\mathrm{PEB}_F$ for both strategies.}

\vspace{-0pt}
\subsubsection{Localization Accuracy Analysis of POLO-II and Comparison with POLO-I}
{We illustrate the RMSE heat map in Fig.~\ref{fig_rmse_heatmap_4AP} for the AP selection strategy proposed in Section~\ref{sec_cov_comp_POLO_II}, using the POLO-II algorithm. We also plot in Fig.~\ref{fig_cdf_4AP} the CDF of the RMSE values obtained from this heat map, and compare it with: (i) the CDF of  $\mathrm{PEB}_F$; and (ii) the CDF of RMSE values corresponding to the heat map generated using AP selection strategy 2 of POLO-I (in Fig. \ref{fig_rmse_col}). We observe that, at this transmit power, POLO-II achieves RMSE values that closely follow  $\mathrm{PEB}_F$ in almost the entire coverage area. In contrast, the RMSE with POLO-I closely follows  $\mathrm{PEB}_F$ in only about $80\%$ of the region.
{This difference is also evident visually: the POLO-I RMSE heat map in Fig.~\ref{fig_rmse_col} exhibits large high-error (yellow) regions, while the POLO-II heat map in Fig.~\ref{fig_rmse_heatmap_4AP} shows very limited high-error areas.}
This clearly demonstrates the superiority of POLO-II over POLO-I.}
\begin{figure}[t]
	\centering
	\begin{subfigure}[b]{\linewidth}\centering
		\includegraphics[width=0.8\linewidth,height=0.7\linewidth]{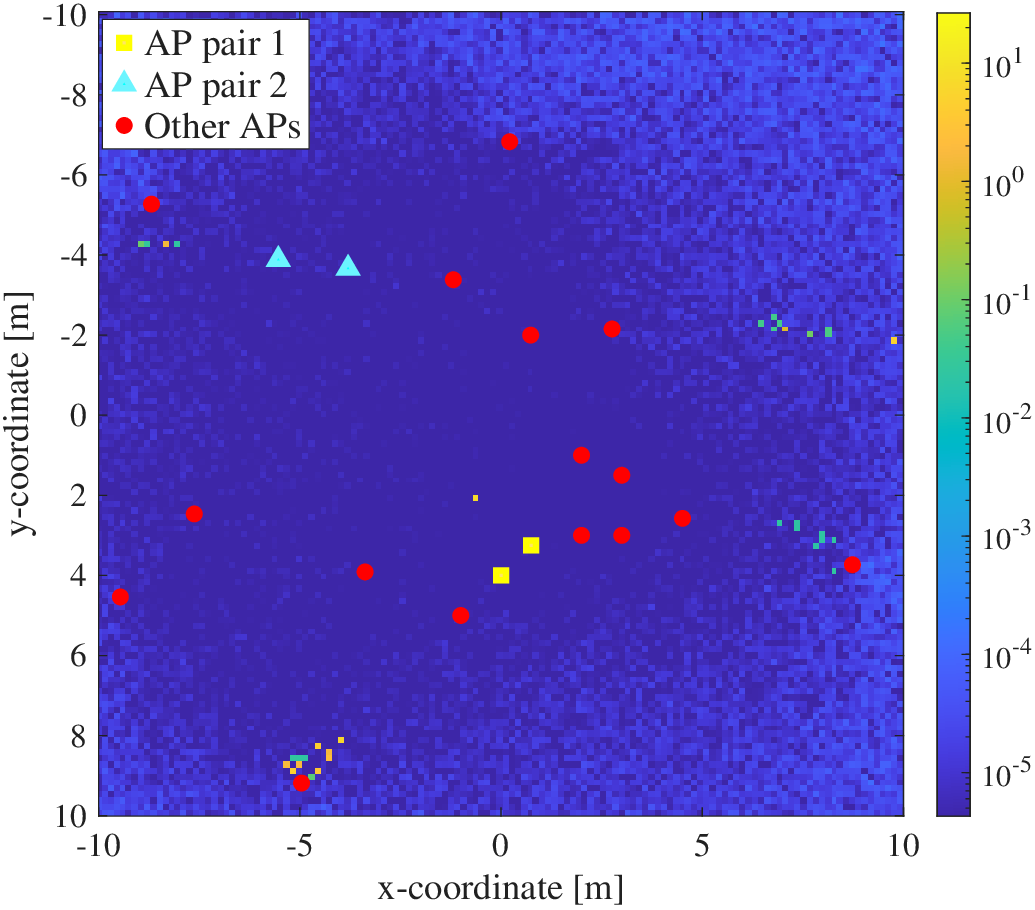}\vspace{0pt}
	\end{subfigure}
	\caption{  RMSE [in m] heat map of POLO-II following proposed strategy in Section~\ref{sec_cov_comp_POLO_II}  for different UE locations in the coverage area.}  
	\label{fig_rmse_heatmap_4AP}\vspace{0pt}
\end{figure}
\begin{figure}[t]
\vspace{-45pt}
    \centering
    \input{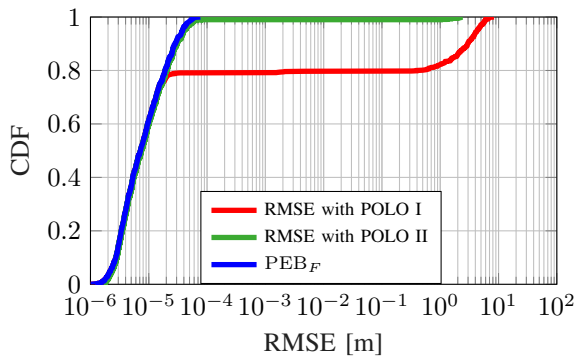}
    \caption{CDF of the RMSE values obtained from the heat map in Fig.~\ref{fig_rmse_heatmap_4AP}, compared with (i) the PEB; and (ii) the RMSE distribution of POLO-I with AP selection strategy 2 (Fig. \ref{fig_rmse_col}).  POLO-II achieves RMSE values close to $\mathrm{PEB}_F$ in $99\%$ of the coverage area, compared to $80\%$ with POLO-I.}
    \label{fig_cdf_4AP}
\end{figure}
\begin{figure*}[t]
	\centering
    \begin{subfigure}[b]{0.32\linewidth}
    \includegraphics[width=\linewidth,height=0.8\linewidth]
    {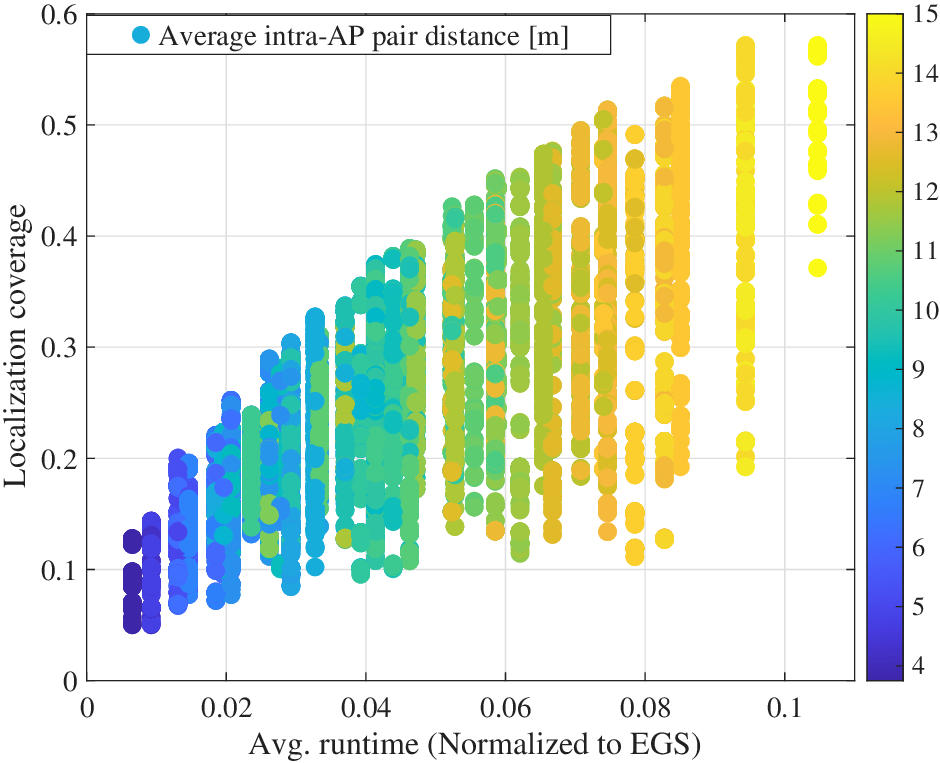}
    \caption{}
    \label{fig_perf_comp_3AP}
    \end{subfigure}
    \begin{subfigure}[b]{0.32\linewidth}
		\includegraphics[width=\linewidth,height=0.8\linewidth]
        {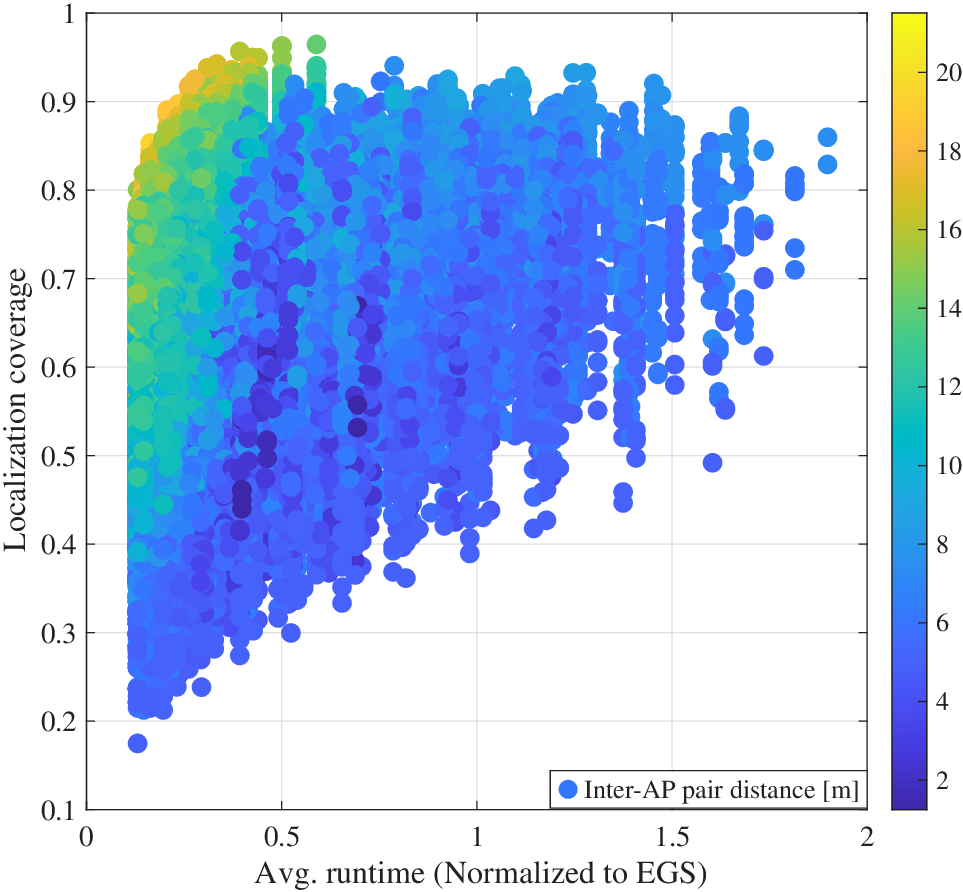}\vspace{0pt}
        \caption{}
        \label{fig_comp_cov_4AP}
	\end{subfigure}  
	\begin{subfigure}[b]{0.32\linewidth}
		\includegraphics[width=\linewidth,height=0.8\linewidth]
        {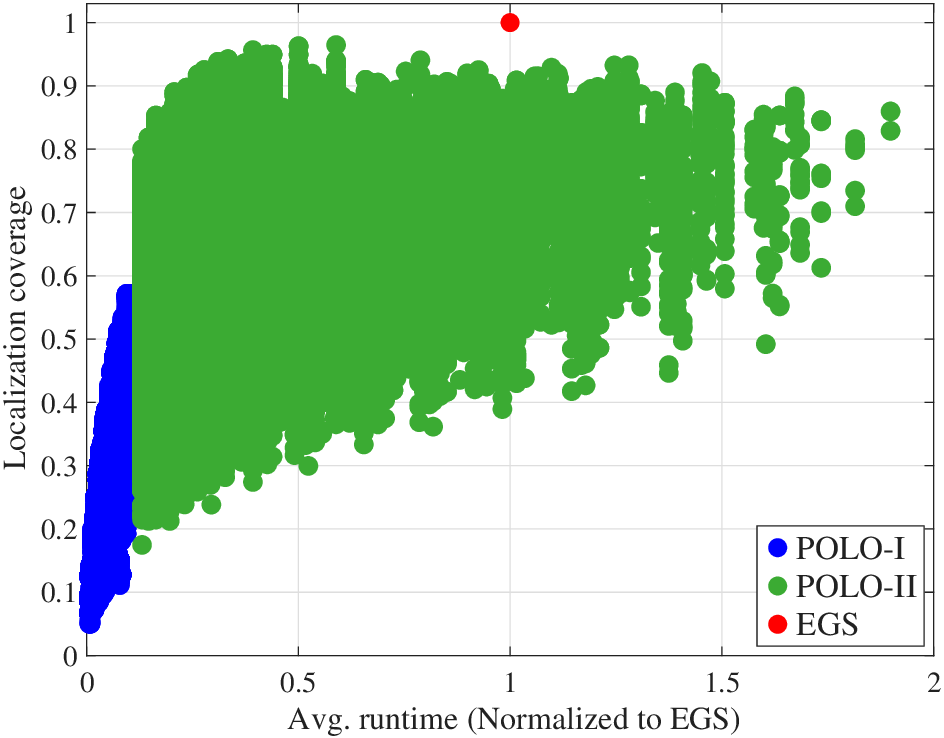}\vspace{0pt} 
        \caption{}
        \label{fig_comp_cov_all}
	\end{subfigure} 
    \caption{{Coverage–complexity tradeoff for (a) POLO-I, (b) POLO-II, and (c) their comparison with EGS. In (a), each point corresponds to an AP triplet, where larger intra-AP distances improve coverage but increase runtime. In (b), each point corresponds to an AP quadruplet, where high coverage with low runtime is achieved by large inter-AP pair distances, illustrating the design principle of POLO-II. Panel (c) combines the data from (a) and (b) and includes EGS for benchmarking, showing POLO-I’s efficiency at low runtime, POLO-II’s superior coverage–runtime balance, and EGS’s ability to achieve full coverage at the cost of high runtime.
 } \vspace{-15pt}} 
	\label{fig_comp_cov}\vspace{-5pt}
\end{figure*}

\subsection{Coverage-Complexity Trade-off Analysis for POLO-I, POLO-II and EGS Algorithms}
{In this subsection, we evaluate the coverage and computational complexity of the proposed POLO-I and POLO-II algorithms under different AP configurations. To do this, the $20$ m$\times20$ m  coverage area is divided into a grid of $2.5$ m$\times2.5$ m squares. Each grid center is treated as a potential AP location, and all possible AP triplets (for POLO-I) and AP quadruplets (for POLO-II) are generated.}

{Our objective is to identify AP configurations that offer the best tradeoff between localization coverage and computational cost. Effectively, this study examines how the spatial arrangement of the APs influences both localization coverage and complexity, providing insights that can guide the intuition on what a formal AP placement optimization would yield.}

\subsubsection{POLO-I}
For each AP triplet, we compute the average intra-AP pair distance by treating each AP once as reference and averaging distances to the remaining two. Among the three possible reference choices, we retain the configuration with the lowest average intra-AP pair distance. Triplets with average intra-AP distances exceeding $15$~m are discarded to limit complexity. 
Each selected triplet is evaluated based on:
\begin{enumerate}
\item \textit{Normalized average runtime with respect to EGS}: Averaged runtime over all UE positions in the coverage area, and then normalized by the runtime of EGS (with grid resolution $0.1\lambda\times0.1\lambda$).
\item \textit{Localization coverage}: Fraction of the coverage area with  $\mathrm{PEB}_T$ (PEB from selected AP triplet) $<\lambda$. In this study, we use  $\mathrm{PEB}_T$ as the performance metric, since, as shown in Section~\ref{sec_AP_sel_2}, it captures how accurately the hyperbola intersection points can pinpoint the minima of the global cost function, and therefore reflects the accuracy of the overall algorithm.
\end{enumerate}
The tradeoff is shown in Fig.~\ref{fig_perf_comp_3AP}, with normalized runtime on the $x$-axis, localization coverage on the $y$-axis, and color indicating average intra-AP distance.

{We observe that larger average intra-AP pair distances lead to both increased coverage and higher runtime.
{This is because, as discussed in Section \ref{sec_AP_sel_2}, the localization performance is best within the segment connecting the selected APs.} Larger AP separation expands this region, improving localization performance across a wider area. This also increases the number of hyperbola intersection points, increasing the computational complexity. This highlights the inherent tradeoff in POLO-I between accuracy and computational cost.}

\subsubsection{POLO-II}
We perform a similar analysis for POLO-II. For each quadruplet, all possible AP pairings are considered, retaining the pairing with the lowest average intra-pair distance. Quadruplets with average intra-pair distance exceeding $15$~m are discarded to limit complexity. Each selected quadruplet is evaluated, as in Fig.~\ref{fig_perf_comp_3AP}, using normalized average runtime (with respect to EGS) and localization coverage, defined as the fraction of the area where the  $\mathrm{PEB}_Q^{\text{ref}}$ (PEB of selected AP quadruplet) is below $\lambda$. Results are shown in Fig.~\ref{fig_comp_cov_4AP}, with normalized runtime in $x$-axis, localization coverage in $y$-axis, and color indicating the inter-AP pair~distance.

{The AP quadruplets of interest in Fig.~\ref{fig_comp_cov_4AP} lie towards the top-left corner. These configurations achieve both low average runtime and  better coverage. We observe that these favorable quadruplets tend to have large inter-AP pair distances. At the same time, they inherently result in low intra-AP pair distances. 
{This is because the APs are deployed within a fixed coverage area, and hence selecting two AP pairs with a larger separation between the pairs inherently results in smaller distances between the APs within each pair.}
These observations are consistent with the insights from Section~\ref{sec_cov_comp_POLO_II}.}

\subsubsection{Comparative Analysis of POLO-I, POLO-II and EGS}
For a clearer comparison between the proposed POLO-I and POLO-II algorithms, along with the benchmark EGS method~\cite{GRID_SEARCH_NAS,MLE_EGS}, we combine the data from Fig. \ref{fig_perf_comp_3AP} and Fig. \ref{fig_comp_cov_4AP} into a single plot in Fig.~\ref{fig_comp_cov_all}. 
{We observe that POLO-I achieves the lowest runtime among all methods, with normalized runtimes below $0.11$ of EGS. This efficiency stems from its use of closed-form solutions for hyperbola intersections. However, its maximum coverage is limited to $55\%$, even at the highest runtime. 
POLO-II, despite lacking closed-form intersection solutions,  can reach over $80\%$ coverage at normalized runtimes just above $0.11$, and up to $95\%$ coverage at around $50\%$ of EGS runtime. This substantial coverage–runtime improvement arises from decoupling coverage from complexity via larger inter-AP pair distances. While computational cost is higher than POLO-I, the superior coverage outweighs this drawback, pushing the Pareto frontier closer to the ideal of maximum coverage and minimum runtime.  
Finally, EGS achieves full ($100\%$) coverage but at the cost of higher computation time.}  
{This comparison highlights the flexibility of POLO-II in balancing performance and complexity, and the efficiency of POLO-I for stringent runtime constraints.}

\section{{Conclusion}}
{In this paper, we introduced POLO, a framework for uplink localization in phase-coherent D-MIMO systems that achieves sub-cm accuracy using single-snapshot, single-subcarrier observations at single-antenna APs. The results show that POLO provides a flexible means to navigate performance–complexity trade-offs, making it suitable for diverse 6G D-MIMO operational requirements. Two POLO variants were proposed: POLO-I, offering real-time feasibility through analytical solutions and serving as a lightweight alternative to ML-based methods; and POLO-II, achieving higher coverage with a modest runtime increase, representing a practical balance between accuracy and complexity. 
A key takeaway is that the AP geometry critically governs both localization accuracy and computational cost. A poorly conditioned AP configuration can lead to unbounded errors over large coverage areas. We proposed different geometry-aware AP selection strategies, which significantly enhances localization coverage while keeping the computational complexity within practical limits. }

Future extensions of POLO may include explicit modeling of NLoS channel to enhance robustness in multipath-dense environments. Advanced AP selection mechanisms can also be considered, that combine observations from multiple AP triplets or quadruplets to construct candidate UE locations. 
Such cooperative selection can help localize the UE across the whole coverage area. If one AP configuration yields unbounded errors, others can still provide valid estimates, thereby ensuring more robust and reliable localization performance.

\appendices 
\section{Closed-Form Hyperbola Intersections Solutions}\label{app_hyp_intersection}

The intersection of hyperbolas can be obtained in closed form by first rewriting \eqref{eq:hyp_1} and \eqref{eq:hyp_2} as~\cite{HYP_SOL_KCHO,TDOA_ANALYT_SOL}:
\begin{align*}
\mathcal{H}_1 &: \|\widetilde{\mathbf{p}}_{s_1} - \widetilde{\mathbf{u}}\| - \|\widetilde{\mathbf{u}}\| = v_1(\widetilde{z}_{s_1}) = v_1, \\
\mathcal{H}_2 &: \|\widetilde{\mathbf{p}}_{s_2} - \widetilde{\mathbf{u}}\| - \|\widetilde{\mathbf{u}}\| = v_2(\widetilde{z}_{s_2}) = v_2,
\end{align*}
where $\widetilde{(\cdot)}$ indicates a relative vector w.r.t. the position of the reference AP, e.g., $\widetilde{\mathbf{p}}_{s_1} =\mathbf{p}_r - \mathbf{p}_{s_1}$. 
The terms $v_1(\widetilde{z}_{s_1}) = \Delta_{s_1} - \widetilde{z}_{s_1} \lambda$ and $v_2(\widetilde{z}_{s_2}) = \Delta_{s_2} - \widetilde{z}_{s_2} \lambda$ depend on the integer ambiguity $\widetilde{z}_{s_1}$ and $\widetilde{z}_{s_2}$, respectively. 
Rearranging the $i$-th hyperbola and squaring both sides, we obtain:
\begin{align}
&\frac{\|\widetilde{\mathbf{p}}_{s_i}\|^2 - v_i^2}{2} = v_i \|\widetilde{\mathbf{u}}\| + \widetilde{\mathbf{p}}_{s_i}^\top \widetilde{\mathbf{u}}. \label{eq:hyp_rearranged}
\end{align}
Collecting the two equations from \eqref{eq:hyp_rearranged} for $i=1,2$, we write the equations in matrix form to get:
\begin{align}
\mathbf{c} = \mathbf{v} \|\widetilde{\mathbf{u}}\| + \mathbf{B} \widetilde{\mathbf{u}},
\end{align}
where $\mathbf{c}=\Big[{(\Vert\widetilde{\mathbf{p}}_{s_1}\Vert^{2}-v_{1}^{2})}/{2},{(\Vert\widetilde{\mathbf{p}}_{s_2}\Vert^{2}-v_{2}^{2})}/{2}\Big]^\top$, $\mathbf{v}=[v_1,v_2]^\top$, and $\mathbf{B}=[\widetilde{\mathbf{p}}_{s_1},\widetilde{\mathbf{p}}_{s_2}]^\top$.
Solving for $\widetilde{\mathbf{u}}$, we obtain:
\begin{align}
\widetilde{\mathbf{u}} = \mathbf{B}^\dagger (\mathbf{c} - \mathbf{v} \|\widetilde{\mathbf{u}}\|) = \bm{\alpha} - \bm{\beta} \|\widetilde{\mathbf{u}}\|, \label{eq:U_norm_U}
\end{align}
where $\mathbf{B}^\dagger$ is the pseudo-inverse of $\mathbf{B}$, $\bm{\alpha} = \mathbf{B}^\dagger \mathbf{c}$ and $\bm{\beta} = \mathbf{B}^\dagger \mathbf{v}$.
Taking the squared norm of both sides of \eqref{eq:U_norm_U}, we~obtain:
\begin{align}
    &\Vert\widetilde{\mathbf{u}}\Vert^2=\Vert\bm{\alpha}-\bm{\beta}\Vert\widetilde{\mathbf{u}}\Vert\Vert^2\nonumber\\
\implies&0=\Vert\bm{\alpha}\Vert^2+2\bm{\alpha}^\top\bm{\beta}\Vert\widetilde{\mathbf{u}}\Vert+(\Vert\bm{\beta}\Vert^2-1)\Vert\widetilde{\mathbf{u}}\Vert^2.\label{eq:quad}
\end{align}
Equation \eqref{eq:quad} is a quadratic in $\|\widetilde{\mathbf{u}}\|$, and can yield 0, 1, or 2 non-negative solutions. Each feasible solution corresponds to a candidate UE position, which can be calculated using \eqref{eq:U_norm_U}.

\section{EFIM for UE Position in POLO-I}\label{app:PEB}
We begin by deriving the fisher information matrix (FIM) for the parameter vector $\bm{\eta}= [\mathbf{u}^\top,\phi,\bm{\rho}^\top]^\top\in\mathbb{R}^{(M+3)\times 1}$. The elements of the FIM $\mathbf{J}$ are given as~\cite[Sec 15.7]{SMKAY_BOOK_1}: 
\begin{align}\label{eq:FIM_full}
        [\mathbf{J}]_{ij} = \frac{2}{\sigma_n^2}\sum_{m=1}^{M}\mathcal{R}\left\{\frac{\partial \mu_m^*}{\partial [\bm{\eta}]_i}\frac{\partial \mu_m}{\partial [\bm{\eta}]_j}\right\},
    \end{align}
where $\mu_m= \sqrt{P}\rho_{m}e^{-\jmath\frac{2\pi }{\lambda} d_{m}} e^{- \jmath \phi} s$ denotes the noise free signal component of $y_m$ in \eqref{eq:signal}. 
The EFIM corresponding to the UE position vector $\mathbf{u}$ is derived by partitioning the parameter vector $\bm{\eta}$ into the parameters of interest—i.e., the UE position $\mathbf{u}$—and the nuisance parameters $\mathbf{w} = [\phi, \bm{\rho}^\top]^\top$. Accordingly, the FIM is block-partitioned as~\cite{RS_UPLINK_JOURNAL}:
\begin{align}
        &\mathbf{J} 
        =
        \begin{bmatrix}
             \mathbf{J}_{uu}
            &\mathbf{J}_{uw}
            \\ 
            \mathbf{J}_{uw}^\top
            &\mathbf{J}_{ww}
        \end{bmatrix},
\end{align}
where $\mathbf{J}_{uu}\in\mathbb{R}^{2\times 2}$ corresponds to the UE position parameters, $\mathbf{J}_{ww}\in\mathbb{R}^{(M+1)\times (M+1)}$ to the nuisance parameters $[\phi,\bm{\rho}^\top]^\top$, and $\mathbf{J}_{uw}\in\mathbb{R}^{2\times (M+1)}$ captures the cross-information between them. The EFIM, obtained via the standard Schur complement, is expressed as $\mathbf{J}_e = \mathbf{J}_{{u}{u}} - \mathbf{J}_{{u}{g}} \mathbf{J}_{{g}{g}}^{-1} \mathbf{J}_{{u}{g}}^\top$~\cite{EFIM_BOOK}. After algebraic simplifications it becomes:
\begin{align}\label{eq:EFIM_final}
    &\mathbf{J}_e
   =\sum_{m=1}^M
        \alpha_m\mathbf{v}_m\mathbf{v}_m^\top
        -\Big(\sum_{m=1}^M \gamma_m\mathbf{v}_{m}\Big)
        \left(\sum_{m=1}^M\gamma_m\mathbf{v}_{m}^\top\right),
\end{align}
where $\alpha_m={8P\pi^2\rho_m^2}/{(\sigma_w^2\lambda^2)}$, and $\gamma^2_m = { {8P\pi^2\rho_{m}^4}/{(\sigma_w^2\lambda^2  \sum\limits_{m'=1}^M\mathcal\rho_{m'}^2)}}$. Here, $\mathbf{v}_m = ({\mathbf{p}_m - \mathbf{u}})/{d_m}$ represents the vector from the UE to the $m$-th AP. 
This EFIM, when evaluated considering only the selected AP triplet, after algebraic manipulations  is rewritten as follows:
\begin{align}
 \mathbf{J}_{T}{(\mathbf{u})} = K_c\sum_{m\in\mathcal{M}}\rho_m^2\mathbf{v}_m\Big(\beta_m \mathbf{v}_m^\top
    -\!\!\!\!\sum_{m'\in\mathcal{M}\backslash m}\rho_{m'}^2\mathbf{v}_{m'}^\top\Big),\!\!\!\!
\end{align}
where  $K_c\!=\! {8P\pi^2}/{(\sigma_n^2\lambda^2\!\!\sum\limits_{m'\in \mathcal{M}}\!\rho_{m'}^2)}$, and $\beta_m\!=\!\sum\limits_{m'\in\mathcal{M}\backslash m}\!\!\rho_{m'}^2$.



\begin{thebibliography}{10}
\providecommand{\url}[1]{#1}
\csname url@samestyle\endcsname
\providecommand{\newblock}{\relax}
\providecommand{\bibinfo}[2]{#2}
\providecommand{\BIBentrySTDinterwordspacing}{\spaceskip=0pt\relax}
\providecommand{\BIBentryALTinterwordstretchfactor}{4}
\providecommand{\BIBentryALTinterwordspacing}{\spaceskip=\fontdimen2\font plus
\BIBentryALTinterwordstretchfactor\fontdimen3\font minus \fontdimen4\font\relax}
\providecommand{\BIBforeignlanguage}[2]{{%
\expandafter\ifx\csname l@#1\endcsname\relax
\typeout{** WARNING: IEEEtran.bst: No hyphenation pattern has been}%
\typeout{** loaded for the language `#1'. Using the pattern for}%
\typeout{** the default language instead.}%
\else
\language=\csname l@#1\endcsname
\fi
#2}}
\providecommand{\BIBdecl}{\relax}
\BIBdecl

\bibitem{LOC_SURVEY_1}
J.~A. del Peral-Rosado \emph{et~al.}, ``Survey of cellular mobile radio localization methods: From {1G} to {5G},'' \emph{{IEEE} Commun. Surveys Tuts.}, vol.~20, no.~2, pp. 1124--1148, 2017.

\bibitem{6g_loc_enabler}
S.~E. Trevlakis \emph{et~al.}, ``Localization as a key enabler of {6G} wireless systems: A comprehensive survey and an outlook,'' \emph{IEEE Open Journal of the Communications Society}, vol.~4, pp. 2733--2801, 2023.

\bibitem{5g_pos_ericsson}
S.~Dwivedi \emph{et~al.}, ``Positioning in {5G} networks,'' \emph{{IEEE} Commun. Mag.}, vol.~59, no.~11, pp. 38--44, 2021.

\bibitem{LOC_TUT_ABU}
M.~Abuyaghi \emph{et~al.}, ``Positioning in {5G} networks: Emerging techniques, use cases, and challenges,'' \emph{{IEEE} Internet Things J.}, vol.~12, no.~2, pp. 1408--1427, 2025.

\bibitem{LOC_SURVEY_5}
X.~Li \emph{et~al.}, ``Massive {MIMO}-based localization and mapping exploiting phase information of multipath components,'' \emph{{IEEE} Trans. Wireless Commun.}, vol.~18, no.~9, pp. 4254--4267, 2019.

\bibitem{NF_SURVEY_1}
J.~An \emph{et~al.}, ``Near-field communications: Research advances, potential, and challenges,'' \emph{{IEEE} Wireless Commun.}, vol.~31, no.~3, pp. 100--107, 2024.

\bibitem{NF_SURVEY_2}
M.~Cui \emph{et~al.}, ``Near-field {MIMO} communications for {6G}: Fundamentals, challenges, potentials, and future directions,'' \emph{{IEEE} Commun. Mag.}, vol.~61, no.~1, pp. 40--46, 2022.

\bibitem{6g_loc_sen_Hui_2024}
H.~Chen \emph{et~al.}, ``{6G} localization and sensing in the near field: Features, opportunities, and challenges,'' \emph{{IEEE} Wireless Commun.}, vol.~31, no.~4, pp. 260--267, 2024.

\bibitem{CPP_EMIL}
J.~Nikonowicz \emph{et~al.}, ``Indoor positioning in {5G}-advanced: Challenges and solution toward centimeter-level accuracy with carrier phase enhancements,'' \emph{{IEEE} Wireless Commun.}, vol.~31, no.~4, pp. 268--275, 2024.

\bibitem{CPP_FAN}
S.~Fan \emph{et~al.}, ``Carrier phase-based synchronization and high-accuracy positioning in {5G} new radio cellular networks,'' \emph{{IEEE} Trans. Commun.}, vol.~70, no.~1, pp. 564--577, 2021.

\bibitem{CPP_HENK}
H.~Wymeersch \emph{et~al.}, ``Fundamental performance bounds for carrier phase positioning in cellular networks,'' in \emph{GLOBECOM 2023-2023 IEEE Global Communications Conference}.\hskip 1em plus 0.5em minus 0.4em\relax IEEE, 2023, pp. 7478--7483.

\bibitem{CPP_TALVITIE}
J.~Talvitie \emph{et~al.}, ``Orientation and location tracking of {XR} devices: {5G} carrier phase-based methods,'' \emph{{IEEE} J. Sel. Topics Signal Process.}, vol.~17, no.~5, pp. 919--934, 2023.

\bibitem{CPP_MIKKO}
F.~Ayten \emph{et~al.}, ``Phase-only positioning: Overcoming integer ambiguity challenge through deep learning,'' \emph{arXiv preprint arXiv:2506.07890}, 2025.

\bibitem{CPP_ZHANG}
Z.~Zhang \emph{et~al.}, ``Indoor carrier phase positioning technology based on {OFDM} system,'' \emph{Sensors}, vol.~21, no.~20, p. 6731, 2021.

\bibitem{CPP_JIN}
C.~Jin \emph{et~al.}, ``A sub-meter accurate positioning using {5G} double-difference carrier phase measurements,'' in \emph{2023 IEEE/ION Position, Location and Navigation Symposium (PLANS)}.\hskip 1em plus 0.5em minus 0.4em\relax IEEE, 2023, pp. 1176--1183.

\bibitem{CPP_KIM}
W.~Kim \emph{et~al.}, ``Implementation of carrier phase positioning for {5G OFDM} system,'' in \emph{2022 13th International Conference on Information and Communication Technology Convergence (ICTC)}.\hskip 1em plus 0.5em minus 0.4em\relax IEEE, 2022, pp. 2058--2061.

\bibitem{CPP_FAN2}
S.~Fan \emph{et~al.}, ``Mobile feature enhanced high-accuracy positioning based on carrier phase and {Bayesian} estimation,'' \emph{{IEEE} Internet Things J.}, vol.~9, no.~16, pp. 15\,312--15\,322, 2022.

\bibitem{CPP_FOUDA}
A.~Fouda \emph{et~al.}, ``Toward cm-level accuracy: Carrier phase positioning for {IIoT} in {5G}-advanced {NR} networks,'' in \emph{2022 IEEE 33rd Annual International Symposium on Personal, Indoor and Mobile Radio Communications (PIMRC)}.\hskip 1em plus 0.5em minus 0.4em\relax IEEE, 2022, pp. 782--787.

\bibitem{DOWNLINK_DIST}
S.~Dey \emph{et~al.}, ``Joint localization and synchronization in downlink distributed {MIMO},'' \emph{arXiv preprint arXiv:2504.10087}, 2025.

\bibitem{RS_UPLINK_JOURNAL}
A.~Fascista \emph{et~al.}, ``Joint localization, synchronization and mapping via phase-coherent distributed arrays,'' \emph{{IEEE} J. Sel. Topics Signal Process.}, vol.~19, no.~2, pp. 412--429, 2025.

\bibitem{vukmirovic2019direct}
N.~Vukmirovi{\'c} \emph{et~al.}, ``Direct wideband coherent localization by distributed antenna arrays,'' \emph{Sensors}, vol.~19, no.~20, p. 4582, 2019.

\bibitem{DPE_GNSS_Magazine_2017}
P.~Closas \emph{et~al.}, ``Direct position estimation of {GNSS} receivers: Analyzing main results, architectures, enhancements, and challenges,'' \emph{{IEEE} Signal Process. Mag.}, vol.~34, no.~5, pp. 72--84, 2017.

\bibitem{vukmirovic2018position}
N.~Vukmirovi{\'c} \emph{et~al.}, ``Position estimation with a millimeter-wave massive {MIMO} system based on distributed steerable phased antenna arrays,'' \emph{EURASIP Journal on Advances in Signal Processing}, vol. 2018, no.~1, p.~33, 2018.

\bibitem{OFDM_DFRC_TSP_2021}
M.~F. Keskin \emph{et~al.}, ``Limited feedforward waveform design for {OFDM} dual-functional radar-communications,'' \emph{{IEEE} Trans. Signal Process.}, vol.~69, pp. 2955--2970, 2021.

\bibitem{Silvia_OFDM_sparse_TWC_2025}
S.~Mura \emph{et~al.}, ``Optimized waveform design for {OFDM}-based {ISAC} systems under limited resource occupancy,'' \emph{{IEEE} Trans. Wireless Commun.}, vol.~24, no.~6, pp. 5241--5254, 2025.

\bibitem{DMIMO_FOR_6G}
O.~Haliloglu \emph{et~al.}, ``Distributed {MIMO} systems for {6G},'' in \emph{2023 Joint European Conference on Networks and Communications \& 6G Summit (EuCNC/6G Summit)}.\hskip 1em plus 0.5em minus 0.4em\relax IEEE, 2023, pp. 156--161.

\bibitem{SMKAY_BOOK_1}
S.~M. Kay, ``Statistical signal processing: estimation theory,'' \emph{Prentice Hall}, vol.~1, pp. Chapter--3, 1993.

\bibitem{MLE_LEE}
H.~Lee \emph{et~al.}, ``Performance comparison of numerical optimization algorithms for rss-toa-based target localization,'' in \emph{2023 IEEE 97th Vehicular Technology Conference (VTC2023-Spring)}.\hskip 1em plus 0.5em minus 0.4em\relax IEEE, 2023, pp. 1--6.

\bibitem{GRID_SEARCH_NAS}
P.~Liashchynskyi \emph{et~al.}, ``Grid search, random search, genetic algorithm: a big comparison for {NAS},'' \emph{arXiv preprint arXiv:1912.06059}, 2019.

\bibitem{MLE_EGS}
Z.~Wang \emph{et~al.}, ``Maximum likelihood direct position determination of multiple sources with channel state information,'' \emph{{IEEE} Trans. Veh. Technol.}, 2025.

\bibitem{PAR_EST_BOOK}
J.~Pfanzagl, ``Mathematical theory of statistics (de {Gruyter Studies in Mathematics 7}),'' 1987.

\bibitem{BOYD_CV_OPT}
S.~P. Boyd \emph{et~al.}, \emph{Convex optimization}.\hskip 1em plus 0.5em minus 0.4em\relax Cambridge university press, 2004.

\bibitem{HYP_SOL_KCHO}
Y.~Chan \emph{et~al.}, ``A simple and efficient estimator for hyperbolic location,'' \emph{{IEEE} Trans. Signal Process.}, vol.~42, no.~8, pp. 1905--1915, 1994.

\bibitem{TDOA_ANALYT_SOL}
P.~Hub{\'a}{\v{c}}ek \emph{et~al.}, ``The complete analytical solution of the {TDOA} localization method,'' \emph{Defence Science Journal}, vol.~72, no.~2, pp. 227--235, 2022.

\bibitem{EFIM_BOOK}
H.~L. Van~Trees, \emph{Optimum array processing: Part IV of detection, estimation, and modulation theory}.\hskip 1em plus 0.5em minus 0.4em\relax John Wiley \& Sons, 2002.

\bibitem{matlab_fsolve}
\BIBentryALTinterwordspacing
{The MathWorks, Inc.}, \emph{Optimization Toolbox User’s Guide (R2024a)}, Natick, MA, USA, 2024, accessed: Nov. 2, 2025. [Online]. Available: \url{https://www.mathworks.com/help/optim/fsolve.html}
\BIBentrySTDinterwordspacing

\bibitem{NUM_OPT}
S.~Wright \emph{et~al.}, ``Numerical optimization,'' \emph{Springer Science}, vol.~35, no. 67-68, p.~7, 1999.

\end{thebibliography}
\end{document}